\documentclass[prc,aps,float,twocolumn,superscriptaddress,showkeys,showpacs,nofootinbib]{revtex4}

\usepackage{amsmath}
\usepackage{amssymb}
\usepackage{amsfonts}
\usepackage{epsfig}
\usepackage{color}

\newcommand{\old}[1]{{\rule{0cm}{0cm}}}

\newcommand{\oldnote}[1]{{\rule{0cm}{0cm}}}

\renewcommand{\vec}[1]{\mathbf{#1}}

\newcommand{\nuc}[2]{{$^{#1}$}{#2}}

\begin{document}

\title{Particle-Number Restoration
       within the Energy Density Functional formalism: \\
       Nonviability of terms depending on noninteger powers
    of the density matrices}

\author{T. Duguet}
\email{thomas.duguet@cea.fr}
\affiliation{National Superconducting Cyclotron Laboratory,
             1 Cyclotron Laboratory,
             East-Lansing, MI 48824, USA}
\affiliation{Department of Physics and Astronomy,
             Michigan State University, East Lansing, MI 48824, USA}
\affiliation{CEA, Centre de Saclay, IRFU/Service de Physique Nucléaire, F-91191 Gif-sur-Yvette, France}

\author{M. Bender}
\email{bender@cenbg.in2p3.fr}
\affiliation{Universit{\'e} Bordeaux,
             Centre d'Etudes Nucl{\'e}aires de Bordeaux Gradignan, UMR5797,
             F-33175 Gradignan, France}
\affiliation{CNRS/IN2P3,
             Centre d'Etudes Nucl{\'e}aires de Bordeaux Gradignan, UMR5797,
             F-33175 Gradignan, France}

\author{K. Bennaceur}
\email{k.bennaceur@ipnl.in2p3.fr}
\affiliation{Universit\'e de Lyon, F-69003 Lyon, France;
Universit\'e Lyon 1, F-69622 Villeurbanne, France; \\
CNRS/IN2P3, UMR 5822, Institut de Physique Nucl\'eaire de Lyon}
\affiliation{CEA, Centre de Saclay, IRFU/Service de Physique Nucléaire, F-91191 Gif-sur-Yvette, France}

\author{D. Lacroix}
\email{lacroix@ganil.fr}
\affiliation{GANIL, CEA et IN2P3, BP 5027, 14076 Caen Cedex, France}

\author{T. Lesinski}
\email{t.lesinski@ipnl.in2p3.fr}
\affiliation{Universit\'e de Lyon, F-69003 Lyon, France;
Universit\'e Lyon 1, F-69622 Villeurbanne, France; \\
CNRS/IN2P3, UMR 5822, Institut de Physique Nucl\'eaire de Lyon}

\date{\today}

\begin{abstract}
We discuss the origin of pathological behaviors that have been
recently identified in particle-number-restoration calculations
performed within the nuclear energy density functional framework.
A regularization method that removes the problematic terms from
the multi-reference energy density functional and which applies (i) to any symmetry restoration- and/or generator-coordinate-method-based
configuration mixing calculation and (ii) to energy
density functionals depending only on integer powers of the
density matrices, was proposed in [D.~Lacroix, T.~Duguet, M.~Bender, arXiv:0809.2041] and implemented for particle-number
restoration calculations in [M.~Bender, T.~Duguet, D.~Lacroix, arXiv:0809.2045]. In
the present paper, we address the viability of non-integer powers
of the density matrices in the nuclear energy density functional.
Our discussion builds upon the analysis already carried out in
[J.~Dobaczewski \emph{et al.}, Phys.\ Rev.\ C \textbf{76}, 054315 (2007)].
First, we propose to reduce the pathological nature of terms depending
on a non-integer power of the density matrices by regularizing the
fraction that relates to the integer part of the exponent using the
method proposed in [D.~Lacroix, T.~Duguet, M.~Bender, arXiv:0809.2041]. Then, we
discuss the spurious features brought about by the remaining fractional
power. Finally, we conclude that non-integer powers
of the density matrices are not viable and should be avoided in the
first place when constructing nuclear energy density functionals that are
eventually meant to be used in multi-reference calculations.
\end{abstract}



\maketitle

\section{Introduction}
\label{intro}

In their recent paper~\cite{doba05a}, Dobaczewski \emph{et al.}\ have pointed out
that there are two distinct pathologies that might appear in calculations aiming at
restoring particle number within the nuclear Energy Density
Functional (EDF) framework. Formulating a Particle Number Restored (PNR) EDF
calculation through a contour integral in the complex plane over multi-reference
(MR) EDF kernels, the two categories of pathologies are associated with spurious
poles and branch cuts of the complex MR-EDF kernels that relate to
dependencies of the latter on integer and non-integer powers of
the (transition) density matrices, respectively.

The possible appearance of spurious poles was already identified in
Refs.~\cite{donau98,anguiano01b,almehed01a}. In Ref.~\cite{lacroix06a}, hereafter
referred to as Paper~I, we demonstrated that such a pathology is shared by any
symmetry-restoration- or
generator-coordinate-method (GCM)-based configuration mixing calculation performed
within the EDF context, which we will call a Multi-Reference Energy Density Functional
(MR-EDF) formalism from hereon. In most other cases than PNR, however, the identification of
the spuriosities is much less transparent. In Paper I, we
proposed a formal and practical regularization method that applies to any symmetry restoration
and/or GCM-based configuration mixing calculation. In Ref.~\cite{bender07x},
hereafter referred to as Paper~II, we applied the correction method to
PNR calculations using a particular energy functional that depends
only on integer powers of the density matrices and thus only display spurious poles.

The pathology associated with spurious branch cuts has been overlooked until
recently~\cite{doba05a} for reasons that will become clear in the following.
As a remedy to it, the authors of Ref.~\cite{doba05a} have proposed to deform
the integration contour in the complex plane such that it does not cross the
branch cuts. As will be discussed below, such a procedure does not allow the
definition of a fully satisfactory theory; e.g.\ the breaking of the shift
invariance remains. In addition, there is no clear method for generalizing the proposed
solution to any other coordinate frequently used in MR-EDF calculations.

In the present paper, we thus address the pathology associated with branch cuts from a different point
of view than in Ref.~\cite{doba05a}. We first make use of the correction
scheme designed in Paper I to regularize the pathology associated with spurious
poles. Doing so we can isolate the part that is specific to the pathology brought
about by branch cuts and question whether it is possible to perform meaningful MR
calculations using an EDF that depends on non-integer powers of the density matrices.
In fact, the question relates to the possibility to deal with any EDF providing
multi-valued MR kernels over the complex plane. It will appear that any EDF (i)
providing multi-valued MR kernels over the complex plane, (ii) whose functional
form is such that the pole structure cannot be extracted analytically; e.g.\ the
family of functionals proposed by Fayans and collaborators~\cite{horen96a,fayans},
is critical. Eventually, anything but low-order polynomials seems difficult, if not
impossible, to handle in practical MR-EDF calculations. Indeed, even if the pole
structure of a complicated EDF can be characterized, it is only for low-order polynomials
that the regularization method proposed in Paper I can be applied to identify the
associated spurious contribution to the physical pole at $z=0$.

The present discussion is conducted for PNR calculations based on a EDF whose
normal part takes the form of a toy Skyrme energy density functional, and whose
pairing part derives from a density-dependent delta interaction (DDDI). Numerical
applications are performed using the realistic SLy4 Skyrme EDF combined with a
local pairing part as derived from a (density-independent) delta interaction (DI).
Two situations of interest are actually considered that correspond to using an
EDF (i) derived from (density-dependent) forces (ii) formulated directly at the
level of the energy functional itself.

The paper is organized as follows. In Sect.~\ref{fracpower}, basic elements
of the single-reference EDF method are recalled and the form of the simplified
energy functional considered for the discussion is given. Section~\ref{PNR}
introduces PNR calculations performed within the EDF framework and describes
the analytical continuation into the complex plane which is used for analysis
purposes in Sect.~\ref{stepsanddivergences}.

Section~\ref{stepsanddivergences}
discusses the occurrence of pathological patterns in particle-number restored energies.
First, we recall the situation for EDFs depending on integer powers of the density
matrices, which is the focus of Papers~I and~II. Then, EDFs depending on non-integer
powers of the density matrices are discussed as the simplest and most practically
relevant example of EDF generating multivalued PNR energy kernels over the complex
plane. Still, the conclusions drawn are valid for more involved EDFs presenting
such a feature. Finally, results of numerical applications are provided in
Sect.~\ref{applications}, highlighting again the differences between EDFs depending
on integer powers of the density matrices and those depending on non-integer ones.
Conclusions are given in Sect.~\ref{conclusions}.

\section{Single-reference EDF method}
\label{fracpower}

Before we present results obtained with a realistic SLy4+DI EDF, we analyze the
relevant physics with a toy functional, reduced to the bare minimum of terms
necessary to convey our point.

\subsection{Density Matrices}

The implementation of the Single-Reference EDF approach relies on the use of a
quasi-particle vacuum $| \Phi_{\varphi} \rangle$ to calculate the one-body
density matrices the energy $\mathcal{E}[\rho,\kappa,\kappa^{\ast}]$ is a functional
of. The index $\varphi$ in $| \Phi_{\varphi} \rangle$ denotes the gauge angle that
provides the orientation of the system in gauge space. Using the requirement that
a meaningful energy functional should be invariant under gauge space rotations, the
angle can be set to a convenient value, usually $\varphi = 0$.

In the canonical basis $\{\phi_{\mu} (\vec{r}) \equiv
\langle \vec{r} | a^{\dagger}_{\mu} | 0 \rangle\}$
of the Bogoliubov transformation that underlies the quasi-particle vacuum
$| \Phi_{0} \rangle$, the SR normal density matrix $\rho$ and anomalous density
matrix $\kappa$ (pairing tensor) take the form
\begin{eqnarray}
\label{intrdens1}
\rho_{\mu \nu}
& \equiv & \frac{\langle \Phi_{0} | a^{\dagger}_{\nu} a_{\mu}
            | \Phi_{0} \rangle}
           {\langle \Phi_{0} | \Phi_{0} \rangle}
  =   v^{2}_{\mu} \,
      \delta_{\mu \nu} \, ,
     \\
\label{intrdens2}
\kappa_{\mu \nu}
& \equiv & \frac{\langle \Phi_{0} | a_{\nu} a_{\mu}
            | \Phi_{0} \rangle}
           {\langle \Phi_{0} | \Phi_{0} \rangle}
  =   u_{\mu} v_{\mu} \,
\delta_{\nu\bar{\mu}} \, ,
       \\
\label{intrdens3}
\kappa^{\ast}_{\mu \nu}
& = & \frac{\langle \Phi_{0} | a^{\dagger}_{\mu} a^{\dagger}_{\nu}
            | \Phi_{0} \rangle}
           {\langle \Phi_{0} | \Phi_{0} \rangle}
  =    u_{\mu} v_{\mu}\,
\delta_{\nu\bar{\mu}} \, ,
\end{eqnarray}
where $\{u_{\mu}, v_{\mu}\}$ are BCS-like occupation numbers such that
$u^{2}_{\mu} + v^{2}_{\mu}=1$, $u_{\mu}=u_{\bar{\mu}}>0$ and
$v_{\mu}=-v_{\bar{\mu}}$. The two canonical states $(\mu,\bar{\mu})$ are the so-called \emph{pair
conjugated} states. Based on an appropriate quantum number, the basis can
be split into a positive half $(\mu>0)$ and a negative half $(\mu<0)$.
When a canonical state $\mu$ belongs to one of these halves, its conjugate
state $\bar{\mu}$ belongs to the other half.

From the point of view of their physical
content, currently used nuclear EDFs can be put under the generic
form~\cite{bender03b}
\begin{equation}
\mathcal{E}[\rho,\kappa,\kappa^{\ast}]
  =     \mathcal{E}_{\text{kin}} [\rho]
      + \mathcal{E}_{\text{norm}}[\rho]
      + \mathcal{E}_{\text{pair}}[\rho,\kappa,\kappa^{\ast}]
\,  ,
\end{equation}
where appear the uncorrelated kinetic energy, the normal and the
pairing contributions, respectively. The
contributions from Coulomb interaction and explicit quantum corrections as the
center of mass correction have been omitted for the sake of using simple notations.
Including them will not modify the arguments given below.
From the point of view of their functional dependence on normal and anomalous
density matrices, the different parts of the functional can be formally written
\begin{eqnarray}
\mathcal{E}_{\text{kin}}[\rho]
& \equiv & \mathcal{E}^{\rho} \, , \\
\mathcal{E}_{\text{norm}}[\rho]
& \equiv & \mathcal{E}^{\rho\rho} + \mathcal{E}^{\rho\rho\rho^{\alpha}} \, , \\
\mathcal{E}_{\text{pair}}[\rho,\kappa,\kappa^{\ast}]
& \equiv &   \mathcal{E}^{\kappa\kappa}
           + \mathcal{E}^{\kappa\kappa\rho^{\gamma}} \, ,
\end{eqnarray}
where the superscripts specify the powers of the normal and anomalous density
matrices that contribute to a given term. The focus of the present work is on
the properties of $\mathcal{E}^{\rho\rho\rho^{\alpha}}$ and
$\mathcal{E}^{\kappa\kappa\rho^{\gamma}}$ when such a nuclear EDF is used
in MR calculations. Note that the Slater approximation which is usually used to tackle the exchange part of the Coulomb contribution to the normal part of the EDF is of the form $\mathcal{E}^{\rho\rho\rho^{\alpha}}$.

For the sake of a transparent discussion, we will perform the analysis for
a toy functional limited to the minimum of ingredients necessary to make the
point. For this purpose, we start from a simplified Skyrme interaction containing
the so-called $t_0$ and $t_3$ terms only, which limits the local densities
entering $\mathcal{E}_{\text{norm}}[\rho]$  to those that do not contain spatial
derivatives~\cite{bender03b}. In addition, and because the validity of the points
made together with the conclusions reached do not critically depend on it, we omit
the isospin degree of freedom and consider one nucleon species only throughout the
discussion. Comments on the additional complexity brought by considering neutrons
and protons are added in Sect.~\ref{isospin}. The generalization to a complete
and realistic Skyrme or Gogny EDF is then straightforward.

\subsection{Local densities}

The local matter and spin densities needed to construct $\mathcal{E}_{\text{norm}}[\rho]$
are given by
\begin{eqnarray}
\label{eq:locrho}
\rho (\vec{r})
& \equiv & \sum_{\mu} \phi^{\dagger}_{\mu} (\vec{r}) \,
                      \phi_{\mu} (\vec{r}) \, \rho_{\mu\mu} \, ,
      \\
\label{eq:locs}
\vec{s} (\vec{r})
& \equiv & \sum_{\mu} \phi^{\dagger}_{\mu} (\vec{r}) \,
           \hat{\boldsymbol{\mathbf\sigma}} \, \phi_{\mu} (\vec{r}) \,
           \rho_{\mu\mu} \, ,
\end{eqnarray}
where $\phi_{\mu} (\vec{r})$ and $\hat{\boldsymbol{\mathbf\sigma}}$
denote a canonical single-particle spinor and the vector of Pauli matrices,
respectively. In addition, one needs the local kinetic density
\begin{eqnarray}
\label{eq:loctau}
\tau (\vec{r})
& \equiv & \sum_{\mu}
           \big[ \vec{\nabla} \phi^{\dagger}_{\mu} (\vec{r}) \big]
           \cdot
           \big[ \vec{\nabla} \phi_{\mu} (\vec{r}) \big] \, \rho_{\mu\mu}
\, ,
\end{eqnarray}
to express the kinetic energy. The three previous local densities can be
put under the form
\begin{equation}
f (\vec{r})
= \sum_{\mu} W^{f}_{\mu \mu} (\vec{r})  \, \rho_{\mu\mu}
\, ,
\end{equation}
where $f \in \{\rho, \vec{s},  \tau \}$ and where the explicit form of
$W^{f}_{\mu \nu} (\vec{r})$ can be easily extracted from
Eqs.~(\ref{eq:locrho}-\ref{eq:loctau}); i.e.
\begin{eqnarray}
\label{eq:Wfunctions}
W^{\rho}_{\mu \nu} (\vec{r})
& \equiv & \phi^{\dagger}_{\mu} (\vec{r}) \, \phi_{\nu} (\vec{r}) \, ,
      \\
\vec{W}^{s}_{\mu \nu} (\vec{r})
& \equiv & \phi^{\dagger}_{\mu} (\vec{r}) \,
           \hat{\boldsymbol{\mathbf\sigma}} \,
           \phi_{\nu} (\vec{r}) \, ,
      \\
W^{\tau}_{\mu \nu} (\vec{r})
& \equiv & \big[\vec{\nabla} \phi^{\dagger}_{\mu} (\vec{r})\big] \cdot
           \big[\vec{\nabla} \phi_{\nu} (\vec{r}) \big] \, .
\end{eqnarray}
The densities entering the pairing part of the EDF are the local pair
densities defined as
\begin{eqnarray}
\label{rhotildelocale}
\tilde{\rho} (\vec{r})
& \equiv & 2 \sum_{\mu>0} W^{\tilde{\rho}}_{\mu\bar{\mu}} (\vec{r}) \,
           \kappa_{\bar\mu \mu} \,  .
\end{eqnarray}
Finally, with the symmetries
of the SR and MR EDF calculations assumed here,
$W^{\tilde{\rho}}_{\mu\bar{\mu}} (\vec{r})$ and
$W^{\bar{\rho} \, *}_{\mu\bar{\mu}}$ are equal and given by the
spin-singlet part of the two-body wave function, defined as
\begin{eqnarray}
\label{spinsingletWF}
W^{\tilde{\rho}}_{\mu\nu}(\vec{r})
= W^{\bar{\rho} \, *}_{\mu\nu}(\vec{r})
& \equiv & \sum_{\sigma = \pm 1} \sigma \, \phi_{\mu} (\vec{r} \sigma) \,
           \phi_{\nu} (\vec{r} -\!\sigma) \\
&  =    &  - W^{\tilde{\rho}}_{\nu\mu}(\vec{r})
   =       - W^{\bar{\rho} \, *}_{\nu\mu}(\vec{r})
\, .
\end{eqnarray}

\subsection{Toy energy density functional}
\label{toyskyrme}

The kinetic energy part of the EDF takes the form
\begin{eqnarray}
\mathcal{E}^{\rho}
& \equiv & \int \! d^3r
           \frac{\hbar^{2}}{2m} \tau (\vec{r}) \, ,
\end{eqnarray}
whereas the normal part derives from a toy Skyrme interaction characterized
by\footnote{One could have considered that the terms multiplying $\rho^2$
and $\vec{s}^2$ in $\mathcal{E}^{\rho\rho\rho^{\alpha}}$ present different
exponents.}
\begin{eqnarray}
\label{bilinearrho}
\mathcal{E}^{\rho\rho}
& \equiv & \int \! d^3r \,
           \big[   A^{\rho\rho} \rho^2 (\vec{r})
                 + A^{ss}   \vec{s}^2  (\vec{r})
           \big] \, ,
           \\
\label{fracrho}
\mathcal{E}^{\rho\rho\rho^{\alpha}}
& \equiv & \int \! d^3r \,
          \big[ A^{\rho\rho\rho^{\alpha}} \rho^2 (\vec{r})
                + A^{ss\rho^{\alpha}}       \vec{s}^2 (\vec{r})
          \big] \, \rho^{\alpha}(\vec{r}) \, .
\end{eqnarray}
Finally, the pairing part of the EDF is given as
\begin{eqnarray}
\label{bilinearkappa}
\mathcal{E}^{\kappa\kappa}
& \equiv & \int \! d^3r \, A^{\tilde{\rho} \tilde{\rho}} \,
           \bar{\rho}^* (\vec{r}) \, \tilde{\rho} (\vec{r}) \, ,
           \\
\label{frackappa}
\mathcal{E}^{\kappa\kappa\rho^{\gamma}}
& \equiv & \int \! d^3r \, A^{\tilde{\rho} \tilde{\rho}\rho^{\gamma}} \,
           \bar{\rho}^* (\vec{r}) \, \tilde{\rho} (\vec{r}) \,
           \rho^{\gamma}(\vec{r}) \, ,
\end{eqnarray}
where the superscripts $ff$ and $fff'$ of the $As$ refer to the local
densities the corresponding term depends on. In addition, one can still
read off those superscripts the powers of normal and anomalous density
matrices that the corresponding term incorporate. Note that no hypothesis
about time-reversal invariance of the system has been made. On the other
hand, we limit ourselves to quasi-particle vacua $| \Phi_{\varphi} \rangle$
with an even number-parity quantum number and thus only discuss explicitly
even-even systems.

The part of the EDF which only depends on the normal density matrix
can be derived from a schematic Skyrme \emph{force}
\begin{eqnarray}
v_{sk} (\vec{R}, \vec{r}_{12})
& = & t_0 \, ( 1 +  x_0 \hat{P}_{\sigma} ) \,  \delta (\vec{r}_{12})
      \nonumber \\
&   & + \frac{t_3}{6} \, ( 1  + x_3 \, \hat{P}_{\sigma} ) \,
        \rho^{\alpha}_{0} (\vec{R}) \,  \delta (\vec{r}_{12})
  \, ,
\end{eqnarray}
where $\vec{R}\equiv(\vec{r}_1 + \vec{r}_2)/2$ and
$\vec{r}_{12} \equiv \vec{r}_1 - \vec{r}_2$, whereas
$\hat{P}_{\sigma} \equiv \frac{1}{2}
( 1 + \vec{\sigma}_1 \cdot \vec{\sigma}_2)$ denotes the spin exchange operator.
Computing the normal part of the EDF as the Hartree and Fock contributions
derived from such an empirical effective vertex, one obtains
\begin{subequations}
\begin{alignat}{4}
A^{\rho\rho}  & =  + \tfrac{1}{4} t_0  ( 1- x_0 ) \, ,
              & \quad
A^{\rho\rho\rho^{\alpha}}  & =  + \tfrac{1}{24} t_{3} ( 1- x_{3} ) \, ,
    \label{coeffs1}          \\
A^{ss}  & =  -\tfrac{1}{4} \, t_0 \,( 1- x_0 ) \, ,
              &
A^{ss\rho^{\alpha}}  & =   -  \tfrac{1}{24} t_{3} ( 1- x_{3} )  \, ,
\label{coeffs2}
\end{alignat}
\end{subequations}
which shows that in this case the four coupling constants entering the EDF
depend on two independent parameters only. However, we will also be interested
in EDFs which are not derived from a Skyrme force and for which the four
coupling constants can be chosen independently. For more complete and realistic
functionals, local gauge invariance imposes constraints between certain
coupling constants~\cite{doba95a}.

The part of the EDF which depends on the anomalous density matrix could be derived from the same Skyrme
force. As one usually focuses on the superfluidity
in the spin-singlet/isospin-triplet channel, one would be led in practice to select only a part of the interaction
from the outset. Furthermore, there exists strong theoretical motivations to explicitly disconnect the part of the
EDF responsible for superfluidity from the part that only depends on the normal density matrix~\cite{henley}.
However, such a decoupling between $\mathcal{E}_{\text{norm}}$ and $\mathcal{E}_{\text{pair}}$ is at the origin of
serious problems encountered in MR-EDF calculations~\cite{doba05a,lacroix06a}. We will come back to that in the
following. For now, one can relate the specific local pairing functional given in
Eqs.~(\ref{bilinearkappa}-\ref{frackappa}) to a DDDI vertex of the form
\begin{equation}
\label{pairingformfactor}
v_{\text{pair}} (\vec{R}, \vec{r})
=  \frac{\tilde{t}_{0}}{2} \, ( 1 - \hat{P}_{\sigma} )
   \left[ 1 - \eta \Big(\frac{\rho_{0}(\vec{R})}{\rho_{sat}}\Big)^{\gamma} \right] \,
   \delta (\vec{r}) \, ,
\end{equation}
where $\rho_{sat} = 0.16$ fm$^{-3}$, which leads to
\begin{equation}
A^{\tilde{\rho} \tilde{\rho}}
= \frac{1}{4} \, \tilde{t}_{0} \, , \qquad
A^{\tilde{\rho} \tilde{\rho}\rho^{\gamma}}
=   - \frac{ \eta}{4\rho^{\gamma}_{c}}  \, \tilde{t}_{0}
 \, .
\end{equation}

Independently of the starting point, a quasi-local pairing EDF must be regularized/renormalized as far as its
ultraviolet divergence is concerned~\cite{bulgac1}.

\section{Particle number restoration}
\label{PNR}

\subsection{Notations}
\label{notations}

As extensively discussed in Ref.~\cite{doba05a} and in Paper~II, Particle Number Restoration (PNR) performed
within the EDF framework relies on calculating the energy of the $N$-particle system through a MR energy functional of the form
\begin{equation}
\label{scalar2}
\mathcal{E}^{N}
\equiv \int_{0}^{2\pi} \! \! \! d\varphi \, \frac{e^{-i\varphi N}}{2\pi \, c^{2}_{N}} \,
       \mathcal{E}[0,\varphi] \,  \langle  \Phi_0 | \Phi_{\varphi} \rangle \, ,
\end{equation}
where
\begin{equation}
\label{weight}
c^{2}_{N}
\equiv \int_{0}^{2\pi} d\varphi \, \frac{e^{-i N \varphi}}{2\pi} \, \langle  \Phi_0 | \Phi_{\varphi} \rangle \, ,
\end{equation}
in such a way that $\mathcal{E}^{N}$ depends only \emph{implicitly} on the (normalized) projected state
\begin{equation}
\label{PWF}
| \Psi^N \rangle
\equiv \frac{\hat{P}^N |\Phi_{0} \rangle}{\langle  \Phi_0 | \hat{P}^N |\Phi_{0} \rangle}
= \int_{0}^{2\pi} d{\varphi} \,
  \frac{e^{ - i \varphi N}}{2\pi \, c_{N}} \, | \Phi_{\varphi} \rangle \, .
\end{equation}
The gauge-space-rotated product states constituting the MR set
of interest read, in their common canonical basis, as
\begin{equation}
\label{eq:transfo_phi}
| \Phi_{\varphi} \rangle
\equiv e^{i \varphi \hat N} \, | \Phi_0 \rangle
  =  \prod_{\mu>0}
     \Big( u_{\mu} + v_{\mu} \, e^{2i\varphi} \, a^+_{\mu} \, a^+_{\bar \mu} \Big)
     | 0 \rangle \, ,
\end{equation}
where $| 0 \rangle$ is the particle vacuum. The above form of
$| \Phi_{\varphi} \rangle$ is convenient to compute the overlap between
a rotated state and the unrotated one
\begin{equation}
\label{overlap}
\langle \Phi_0  | \Phi_{\varphi} \rangle
= \prod_{\mu > 0} \big( u_\mu^2 + v_{\mu}^2 e^{2i\varphi} \big)
   \, .
\end{equation}
In Eq.~(\ref{scalar2}), $\mathcal{E}[0,\varphi]$ denotes the (set of) MR
energy density functional kernel(s). It is traditionally defined
by replacing the SR normal and anomalous density matrices by transition ones
\begin{eqnarray}
\label{contractph}
\rho^{0\varphi}_{\mu \nu}
& \equiv & \frac{\langle \Phi_{0} | a^{\dagger}_{\nu} a_{\mu} | \Phi_{\varphi} \rangle}
                {\langle \Phi_{0} | \Phi_{\varphi} \rangle}
  =  \frac{v_{\mu}^2 \, e^{2 i \varphi} }
          {u_\mu^2 + v_{\mu}^2 \, e^{2 i \varphi} } \, \delta_{\nu \mu} \, ,
     \\
\label{contracthh}
\kappa^{0\varphi}_{\mu \nu}
& \equiv & \frac{\langle \Phi_{0} | a_{\nu} a_{\mu}| \Phi_{\varphi} \rangle}
                {\langle \Phi_{0} | \Phi_{\varphi} \rangle}
  =  \frac{u_\mu v_{\mu} \, e^{2 i \varphi} }
          {u_\mu^2 + v_{\mu}^2 \, e^{2 i \varphi} } \, \delta_{\nu \bar{\mu}} \, ,
     \\
\label{contractpp}
\kappa^{\varphi 0 \, \ast}_{\mu \nu}
& \equiv & \frac{\langle \Phi_{0} | a^{\dagger}_{\mu} a^{\dagger}_{\nu} | \Phi_{\varphi} \rangle}
                {\langle \Phi_{0} | \Phi_{\varphi} \rangle}
  =  \frac{u_\mu v_{\mu}}
          {u_\mu^2 + v_{\mu}^2 \, e^{2 i \varphi} } \, \delta_{\nu \bar{\mu}} \, ,
\end{eqnarray}
into the SR EDF $\mathcal{E}[\rho,\kappa,\kappa^{\ast}]$. This corresponds to defining non-diagonal energy
kernels through the prescription
\begin{equation}
\label{eq:kernelsdef}
\mathcal{E}[0,\varphi]
\equiv \mathcal{E}[\rho^{0\varphi},\kappa^{0\varphi},\kappa^{\varphi 0 \, \ast}]
\, .
\end{equation}
As discussed in Paper~I, MR-EDF calculations performed along the lines presented above fulfill basic requirements~\cite{Robledo07a} but may display pathologies such as divergences and finite steps in the energy. The extent of such problems depends on the analytical form of the EDF
used. In order to conduct an in-depth analysis of the potential problems, it is necessary to perform an
analytical continuation of $\mathcal{E}[0,\varphi]$ to the complex plane~\cite{Bay60a,doba05a}.

\subsection{Continuation to the complex plane}

The continuation is achieved by extending the complex number $z=e^{i \varphi}$ onto the entire complex plane in all
previous formulae.\footnote{The same notation as before is used when extending the definition of SR states and energy kernels to any value of the complex variable $z$. Thus, we abusively replace the gauge angle $\varphi$ by the complex variable $z$ in all our expressions; i.e.\ SR states characterized by the gauge angle $\varphi$, $| \Phi_{\varphi} \rangle$ are extended into $| \Phi_{z} \rangle$ to denote SR states anywhere on the complex plane. In particular, the unrotated SR state, denoted as $| \Phi_0 \rangle$ when using $\varphi$ as a variable, is written as $| \Phi_1 \rangle$ when using $z$ as a more general variable.} In that context, the PNR energy defined through Eq.~(\ref{scalar2}) results from integrating over
over a closed contour around $z=0$ which can be chosen as the unit circle $C_{1} \, (|z|=R=1)$
\begin{eqnarray}
\label{projenergy3}
\mathcal{E}^{N}
& \equiv & \oint_{C_{1}} \frac{dz}{2i\pi c^{2}_{N}} \,
           \frac{\mathcal{E} \left[z\right]}{z^{N+1}} \,
           \langle \Phi_1 | \Phi_{z} \rangle \, ,
       \\
\label{denominator2}
c^{2}_{N}
& = & \oint_{C_{1}} \frac{dz}{2i\pi} \,
      \frac{1}{z^{N+1}} \, \langle  \Phi_1 | \Phi_{z} \rangle \, ,
\end{eqnarray}
where
\begin{equation}
\langle  \Phi_1 | \Phi_{z} \rangle
= \prod_{\mu >0} \left(u_{\mu}^2 + v_{\mu}^2 \, z^{2}\right) \, .
\end{equation}
With this continuation, the transition density matrix and pairing tensor
read as
\begin{eqnarray}
\label{contractphcomplex}
\rho^{1z}_{\mu \nu}
& = & \frac{v_{\mu}^2 \, z^{2}}{u_\mu^2 + v_{\mu}^2 \, z^{2}} \,
      \delta_{\nu \mu} \, ,
      \\
\label{contractppcomplex}
\kappa_{\mu \nu}^{1z}
& = & \frac{u_\mu v_{\mu} \, z^{2}}{u_\mu^2 + v_{\mu}^2 \, z^{2}} \,
      \delta_{\nu \bar{\mu}}  \, ,
      \\
\label{contracthhcomplex}
\kappa_{\mu \nu}^{z1 \, \ast}
& = & \frac{u_\mu v_{\mu}}{u_\mu^2 + v_{\mu}^2 \, z^{2}}
       \, \delta_{\nu \bar{\mu}} \, ,
\end{eqnarray}
and must replace the SR density matrices in Eqs.~\ref{eq:locrho}-\ref{rhotildelocale} in order to define the corresponding transition local densities. Finally, the energy kernel from Eq.~(\ref{eq:kernelsdef}) reads as
\begin{equation}
\label{eq:kernelsdef2}
\mathcal{E}[z]
\equiv \mathcal{E}[\rho^{1z},\kappa^{1z},\kappa^{z1\, \ast}]
\, .
\end{equation}

\section{Steps and divergences}
\label{stepsanddivergences}

\subsection{General considerations}

\begin{figure}[t!]
\includegraphics[height=10.cm,angle=-90]{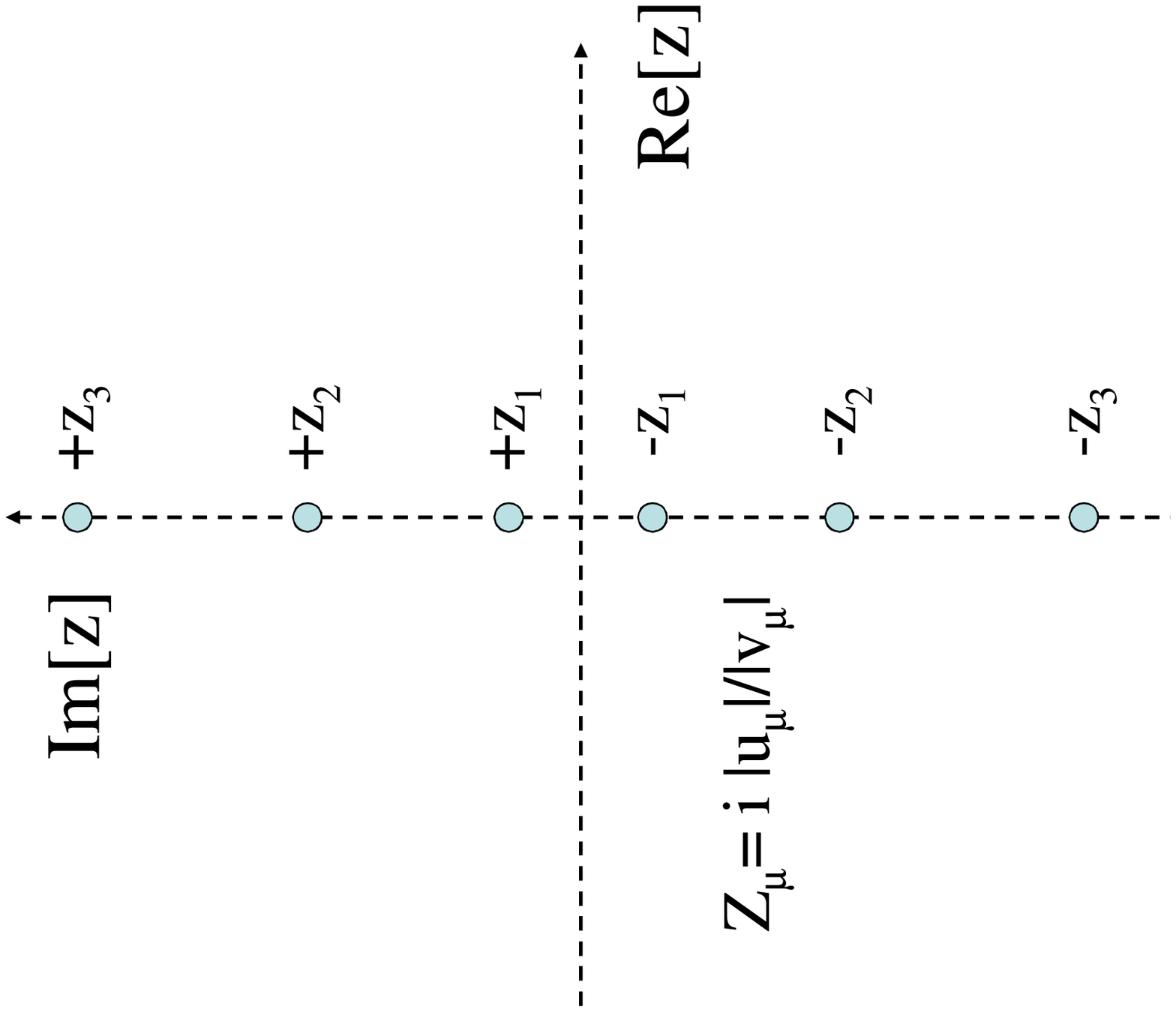}
\caption{\label{poles} Pole structure of  $\rho^{1z}$, $\kappa^{1z}$
and $\kappa^{z1 \, \ast}$ on the complex plane.
}
\end{figure}

The computation of $\mathcal{E}^{N}$ through an integration over a contour encircling
the origin requires the knowledge of the (non-)analytical structure of the integrand
$\mathcal{E}[z] \, \langle \Phi_1 | \Phi_{z} \rangle/z^{N+1}$ over the complex plane.
First, it obviously contains a (physical) pole at $z=0$. Since $\mathcal{E}[z]$ is a
functional of the transition density matrices, (i) it is a function of $z^{2}$ and is
thus even, i.e.\ $\mathcal{E}[z]=\mathcal{E}[-z]$, (ii) its analytical structure
relates to the one of the transition densities. As displayed in Fig.~\ref{poles}, it
is trivial to see that $\rho^{1z}$, $\kappa^{1z}$ and $\kappa^{z1 \, \ast}$ possess
simple poles at $z=\pm z_{\mu} \equiv \pm i |u_{\mu}|/|v_{\mu}|$~\cite{doba05a}. In
general, it is likely that those poles will translate into non-analytical features of
$\mathcal{E}[z] \, \langle \Phi_1 | \Phi_{z} \rangle$ that have serious consequences
on the PNR energy.

As explained in Paper~I, it is necessary to go to configuration space to isolate the
spurious contributions to the MR-EDF energy. For a given pair of vacua belonging to the
MR set, the basis relevant to the analysis of the corresponding energy kernel is the
canonical basis of the Bogoliubov transformation connecting the two vacua. For PNR
calculations, this simply amounts to expressing the EDF kernel $\mathcal{E}[z]$ in the
canonical basis of the Bogoliubov transformation defining any of the product states of
reference; e.g.\ $| \Phi_{1} \rangle$. Indeed, the same canonical basis is shared by all
product states $| \Phi_{z} \rangle$ over the complex plane, as well as by the Bogoliubov
transformation linking any pair of them.

\subsection{Term depending on integer powers of densities}
\label{correctbilinear}

Let us start the analysis with terms that depend on integer powers of the
density matrices. To illustrate the situation, we make use of the bilinear
parts, Eqs.~(\ref{bilinearrho}) and~(\ref{bilinearkappa}), of the toy EDF
introduced in Sect.~\ref{toyskyrme}.

\subsubsection{Matrix elements}

Working in the canonical basis of Bogoliubov transformation connecting
$| \Phi_{1} \rangle$ and $| \Phi_{z} \rangle$, the bilinear part of the
energy kernel $\mathcal{E}[z]$  takes the form
\begin{eqnarray}
\lefteqn{
\mathcal{E}^{\rho\rho}[z] + \mathcal{E}^{\kappa\kappa}[z]
} \nonumber \\
& = & \tfrac{1}{2} \sum_{\mu\nu}
      \bar{v}^{\rho\rho}_{\mu\nu\mu\nu}\,
      \rho^{1z}_{\mu\mu} \, \rho^{1z}_{\nu\nu}
      + \tfrac{1}{4} \sum_{\mu\nu} \bar{v}^{\kappa\kappa}_{\mu\bar{\mu}\nu\bar{\nu}} \,
       \kappa^{z1\, \ast}_{\mu\bar{\mu}} \, \kappa^{1z}_{\nu\bar{\nu}}
 \label{extwick2} \, ,
\end{eqnarray}
where $\bar{v}^{\rho\rho}$ and $\bar{v}^{\kappa\kappa}$ denote matrix elements
of \emph{effective} two-body vertices associated with $\mathcal{E}^{\rho\rho}$
and $\mathcal{E}^{\kappa\kappa}$, respectively. For the toy functional of
Eq.~(\ref{bilinearrho}), the matrix elements of $\bar{v}^{\rho\rho}$ take
the form
\begin{widetext}
\begin{eqnarray}
\label{phmatrixelement:skyrme}
\bar v^{\rho\rho}_{\mu\nu\mu\nu}
& \equiv & 2 \int \! d^3r \;
             \big[ A^{\rho\rho} \, W^{\rho}_{\mu \mu} (\vec{r}) \,
                                   W^{\rho}_{\nu \nu} (\vec{r})
                 + A^{ss} \, \vec{W}^{s}_{\mu \mu} (\vec{r})
                   \cdot \vec{W}^{s}_{\nu \nu} (\vec{r})
             \big] \,
\, .
\end{eqnarray}
\end{widetext}
The quasi-local nature of the Skyrme energy functional (the toy functional
considered  here being purely local) simplifies the construction of the
matrix elements  $\bar v^{\rho\rho}_{\mu\nu\mu\nu}$ as they involve a
single spatial integral only. However, the discussion conducted in the
rest of the paper would hold equally for non-local functionals;
e.g.\ as obtained from finite-range, possibly non-local, effective vertices.

The matrix elements associated with $\mathcal{E}^{\kappa\kappa}$ in
Eq.~(\ref{bilinearkappa}) take the form
\begin{equation}
\label{ppmatrixelement}
\bar v^{\kappa\kappa}_{\mu\bar{\mu}\nu\bar{\nu}}
\equiv  4 \int \! d^3 r \; A^{\tilde{\rho}\tilde{\rho}} \,
                   W^{\bar{\rho} \ast}_{\mu\bar{\mu}} (\vec{r}) \,
                   W^{\tilde{\rho}}_{\nu \bar{\nu}} (\vec{r}) \,
\, .
\end{equation}
Note that for PNR calculations, the matrix elements that one naturally
associate to any term of the EDF depending on integer powers of the
density matrices do not depend on the pair of vacua $| \Phi_{1} \rangle$
and $| \Phi_{z} \rangle$ under consideration, i.e.\ they do not depend
on the gauge variable~$z$.

\subsubsection{Analytical structure of $\left(\mathcal{E}^{\rho\rho}[z]
+ \mathcal{E}^{\kappa\kappa}[z]\right) \,  \langle  \Phi_1 | \Phi_{z} \rangle$}

Due to the additional presence of the norm factor $\langle \Phi_1 |
\Phi_{z} \rangle$ in the integrand of Eq.~(\ref{projenergy3}), it is
easy to realize that only the terms corresponding to $\nu=\mu$ and
$\nu=\bar{\mu}$ in Eq.~(\ref{extwick2}) can lead to non-analytical
features~\cite{lacroix06a,bender07x}. Such terms contribute to the
integrand through
\begin{widetext}
\begin{eqnarray}
\nu=\mu &\Longrightarrow& \tfrac{1}{2} \, \Big( \bar{v}^{\rho\rho}_{\mu \mu \mu \mu} +
\bar{v}^{\rho\rho}_{\bar{\mu} \bar{\mu} \bar{\mu} \bar{\mu}}\Big) \, \frac{v^{4}_{\mu} \, z^{4}}{u_\mu^2 +
v_{\mu}^2 \, z^{2}}  \, \prod_{\nu \neq \mu >0} \left(u_{\nu}^2 + v_{\nu}^2 \, z^{2}\right)
\, , \label{onefermionPNR}\\
\label{enerpairproj1} \nu=\bar{\mu} &\Longrightarrow& \Big[ \tfrac{1}{2}
         \Big( \bar{v}^{\rho\rho}_{\mu \bar{\mu} \mu \bar{\mu}}
               +\bar{v}^{\rho\rho}_{\bar{\mu} \mu \bar{\mu} \mu}
         \Big) \, v^{2}_{\mu} \,  z^{2}
        +\bar{v}^{\kappa\kappa}_{\mu\bar{\mu} \mu \bar{\mu}} \, u^{2}_{\mu}
  \Big] \,
  \frac{v^{2}_{\mu} \,  z^{2}}
       {u_\mu^2 + v_{\mu}^2 \,
  z^{2}} \,
  \prod_{\nu \neq \mu >0} \left(u_{\nu}^2 + v_{\nu}^2 \, z^{2}\right)
\, ,
\end{eqnarray}
\end{widetext}
and both contain potential poles at $z=\pm z_{\mu} = \pm i |u_{\mu}|/|v_{\mu}|$.
Note that those poles do not exist in the first place if the states $(\mu,\bar{\mu})$
are more than doubly degenerate in terms of occupation numbers as an additional factor
from the norm then compensates the single pole in
Eqs.~(\ref{onefermionPNR}-\ref{enerpairproj1}).\footnote{This holds for bilinear
functionals. A term of order $n$ in the density matrices can generate a pole
at $\pm z_{\mu}$ of order (at most) $(n-1)$. For the pole to disappear,
$(n-1)$ additional factors from the norm kernel are needed to cancel the denominator
$(u^{2}_{\mu}+ v^{2}_{\mu} \, z^{2})^{-(n-1)}$. Thus, the pair of interest
($\mu$, $\bar{\mu}$) needs to be degenerate (at least) with $(n-1)$ other pairs
in terms of occupations for this to occur.}

Otherwise, the poles disappear in Eq.~(\ref{onefermionPNR}) if, and only if,
$\bar{v}^{\rho\rho}_{\mu \mu \mu \mu} =
\bar{v}^{\rho\rho}_{\bar{\mu} \bar{\mu} \bar{\mu} \bar{\mu}}=0$; i.e.\ the
matrix elements associated with $\mathcal{E}^{\rho\rho}$ are antisymmetrized.
Coming back to the toy Skyrme functional used in the present paper, and
noticing that
\begin{equation}
|\vec{W}^{s}_{\mu \mu} (\vec{r})|^{2}
=  |W^{\rho}_{\mu \mu} (\vec{r})|^{2}
= \Big[\sum_{\sigma = \pm 1}|\varphi_{\mu}(\vec{r}\sigma)|^{2}\Big]^{2} \, ,
\end{equation}
for all $\mu$, one finds that $\bar{v}^{\rho\rho}_{\mu \mu \mu \mu} =
\bar{v}^{\rho\rho}_{\bar{\mu} \bar{\mu} \bar{\mu} \bar{\mu}}=0$ if, and only if,
$A^{ss} = - A^{\rho\rho}$. As shown by Eqs.~(\ref{coeffs1}-\ref{coeffs2}), such a
condition is satisfied when starting from the (density-independent part of the)
Skyrme \emph{force}. The previous analysis is trivially extended to the
density-independent part of a more complete Skyrme or Gogny vertex. On the other
hand, using a functional approach that bypasses the introduction of a two-body vertex,
relationships such as $A^{ss} = - A^{\rho\rho}$ might not be fulfilled. In such a
case $\mathcal{E}^{\rho\rho}$ generates poles at $z=\pm z_{\mu}$ in the integrand
of Eq.~(\ref{projenergy3}).

The poles disappear from Eq.~(\ref{enerpairproj1}) if, and only if,
$\bar{v}^{\rho\rho}_{\mu \bar{\mu} \mu \bar{\mu}}
= \bar{v}^{\kappa\kappa}_{\mu \bar{\mu} \mu \bar{\mu}}$; i.e.\ diagonal matrix
elements involving two conjugated canonical states are identical in
$\mathcal{E}^{\rho\rho}$ and $\mathcal{E}^{\kappa\kappa}$. If it is so, the two
terms in the bracket of Eq.~(\ref{enerpairproj1}) combine in such a way that the
dangerous denominator explicitly cancels out. One is then left with a finite
contribution to the MR energy kernel. Such a recombination is obviously satisfied
if both $\mathcal{E}^{\rho\rho}$ and $\mathcal{E}^{\kappa\kappa}$ are constructed
from the same (effective) force, for example when using the density-independent
part of the Gogny interaction~\cite{anguiano01b}. Using a functional approach or
starting from two different effective vertices to build $\mathcal{E}^{\rho\rho}$
and $\mathcal{E}^{\kappa\kappa}$, the recombination is unlikely to occur and one
is left with an ill-defined PNR formalism and compromised results. Just as we did
to ensure that $\bar{v}^{\rho\rho}_{\mu \mu \mu \mu} =
\bar{v}^{\rho\rho}_{\bar{\mu} \bar{\mu} \bar{\mu} \bar{\mu}}=0$, i.e.\
$A^{ss} = - A^{\rho\rho}$, one could work out minimal constraints between
the coupling constants entering $\mathcal{E}^{\rho\rho}$ and
$\mathcal{E}^{\kappa\kappa}$ to impose that
$\bar{v}^{\rho\rho}_{\mu \bar{\mu} \mu \bar{\mu}} =
\bar{v}^{\kappa\kappa}_{\mu \bar{\mu} \mu \bar{\mu}}$ in the underlying EDF.

\subsubsection{Projected energy from a Hamiltonian}
\label{projEintegerpowers}

\begin{figure}[t]
\includegraphics[height=10.cm,angle=-90]{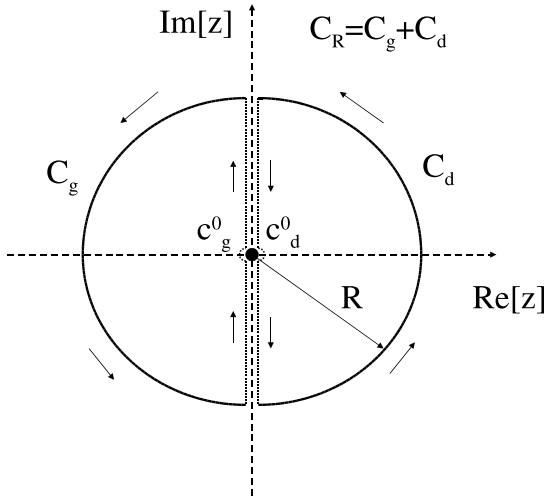}
\caption{\label{wellbehavedintegration2}
Computation of $\mathcal{E}^{N}$ for an EDF (i) obtained from the average
value of a genuine Hamiltonian in the projected state (ii) depending only on
integer powers of the densities and after applying the correction proposed
in Paper~I. The integration is performed in the complex plane over a
circular contour $C_{R}$ of arbitrary radius $R$.
}
\end{figure}

As seen from the previous discussion, poles in the transition densities do not
always translate into poles in $\mathcal{E}[z] \, \langle \Phi_1 | \Phi_{z} \rangle$.
The most trivial example for this occurs when the particle number projected energy
is computed from the average value of a genuine Hamiltonian in the projected state
$| \Psi^{N} \rangle$; i.e.\ what we denote as the strict projected HFB approach
in Paper~II. In this case, the only pole of the integrand in Eq.~(\ref{projenergy3})
is the physical one at $z=0$.  To apply the Cauchy theorem\footnote{The present
Section reformulates parts of the analysis proposed in Paper~II for functionals
proportional to integer powers of the density matrices, i.e.\ we employ Cauchy's
integral theorem rather than using directly Cauchy's residue formula. Coming back
to Cauchy's integral theorem will be needed to conduct the discussion for more
general functionals as is indicated in the next Section.} and calculate the
projected energy, the original circular contour $\mathcal{C}_{1}$ must be deformed
to exclude the pole at $z=0$. As shown in Fig.~\ref{wellbehavedintegration2}, this
can be achieved by choosing two semi-circular contours $C_{d}$ and $C_{g}$, such
that $C_{1}\equiv \big[C_{g}+C_{d}\big](\epsilon \rightarrow 0)$, and by closing
those semi-circular contours along the imaginary axis in such a way that the pole
at $z=0$ is bypassed by two semi-circles of infinitely small radii. Using such
contours, it is easy to prove that
\begin{eqnarray}
\label{projenergy4}
c^{2}_{N} \, \mathcal{E}^{N}
& = &  \mathcal{R}es \left.\left[\frac{\mathcal{E}[z] \, \langle \Phi_1 | \Phi_{z} \rangle}
                                      {z^{N+1}}
                           \right]\right|_{z=0} \, ,   \\
\label{denominator3}
c^{2}_{N}
& = & \mathcal{R}es \left.\left[
      \frac{\langle \Phi_1 | \Phi_{z} \rangle }{z^{N+1}} \right]\right|_{z=0} \, .
\end{eqnarray}
Because the only pole of the integrand is at $z=0$, the same result is obtained
for $\mathcal{E}^{N}$ by starting from any integration contour encircling the
origin in Eq.~(\ref{projenergy3}). When the energy is calculated as the average
value of a Hamiltonian in the projected state, the independence of the projected
energy on the details of the integration contour, as for example its radius, can be
related to the invariance of the normalized projected state with respect to
\emph{shift transformations}~\cite{Bay60a,doba05a}. This symmetry will be discussed
below in the EDF context.

\subsubsection{PNR energy from an EDF}
\label{projEintegerpowersb}

\begin{figure}[t!]
\includegraphics[height=10.cm,angle=-90]{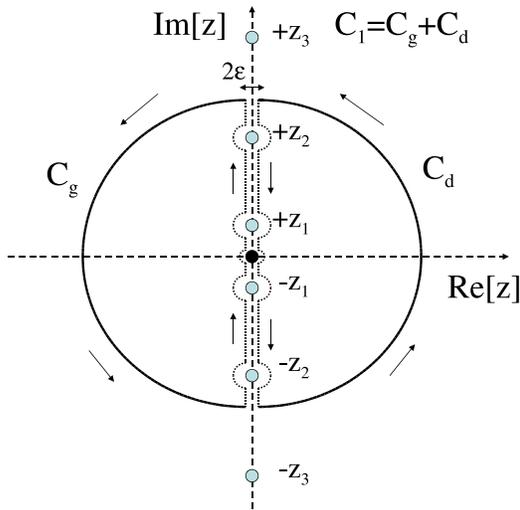}
\caption{\label{wellbehavedintegration}
Computation of $\mathcal{E}^{N}$ for an EDF depending on integer powers of
the densities. The integration is performed in the complex plane over the unit circle $C_{1}$.}
\end{figure}

The poles subsist in Eqs.~(\ref{onefermionPNR}) and~(\ref{enerpairproj1}) for any EDF
that is characterized by $\bar{v}^{\rho\rho}_{\mu \mu \mu \mu} \neq 0$ and/or
$\bar{v}^{\rho\rho}_{\mu \bar{\mu} \mu \bar{\mu}} \neq
\bar{v}^{\kappa\kappa}_{\mu \bar{\mu} \mu \bar{\mu}}$. To apply the Cauchy theorem
in this case, the circular contour $C_{1}$ must now be deformed to exclude not only
the pole at $z=0$ but also those at $z=\pm z_{\mu}$ which are inside the unit circle.
As shown in Fig.~\ref{wellbehavedintegration}, this can be done by choosing two
semi-circular contours $C_{d}$ and $C_{g}$, such that $C_{1}\equiv \big[C_{g} +
C_{d}\big](\epsilon \rightarrow 0)$, and by closing each of them along the imaginary
axis in such a way that all the poles are bypassed by semi-circles of infinitely
small radii. Using such contours, the Cauchy theorem leads to
\begin{eqnarray}
\label{projenergy5}
c^{2}_{N} \, \mathcal{E}^{N}
& = & \sum_{z=0, \pm z_{\mu}} \mathcal{R}es \left.\left[
      \frac{\mathcal{E}[z]  \, \langle \Phi_1 | \Phi_{z} \rangle}{z^{N+1}}\right]\right|_{z} \, ,
\end{eqnarray}
whereas $c^{2}_{N}$ remains unchanged.

According to Eq.~(\ref{projenergy5}), the existence of poles at $z= \pm z_{\mu}$
in the integrand makes the PNR energy to (i) depend on the radius of the integration
circle~\cite{doba05a,bender07x} (ii) display a finite step whenever a pole leaves
the integration circle; e.g. as the system is deformed along a collective degree of
freedom~\cite{doba05a,bender07x}. Such a behavior make the PNR energy to break shift
invariance. This is very undesirable as the concept of shift transformation and shift
invariance can be extended to the EDF framework in such a way that the invariance of
$\mathcal{E}^{N}$ with respect to the radius of the integration contour remains a
fundamental feature of the theory~\cite{duguet06b}.

Also, PNR energies may display divergences whenever a pole crosses the integration
circle. When a pole sits on the integration contour $C_{R}$, the definition of the
contour $C_{R}\equiv\big[C_{g}+C_{d}\big](\epsilon \rightarrow 0)$ is in fact ambiguous
and requires an additional prescription. The most natural procedure is to define the
integration through the pole in the sense of the Cauchy principal value. Doing so
provides a finite PNR energy if the Laurent series of the integrand centered at the
pole only contains odd powers. Considering the structure of the nuclear EDF, this
will happen if the EDF (i) only contains bilinear terms (ii) contains additional
trilinear terms that do not allow three powers of the same isospin (as a zero-range
three-body force does not allow) (iii) contains additional quartic terms which are
bilinear in each isospin. In this case, one is left with simple poles at
$z=\pm z_{\mu}$ and the Cauchy principal value equals half the result that would be
obtained if the pole were to lie inside the integration circle. In all other cases,
one can see that (i) the poles at $z=\pm z_{\mu}$ will be of higher orders (ii) the
Laurent series centered at those poles will contain even powers (ii) the Cauchy
principle value will lead to an infinite values and the PNR energy will diverge as
a poles crosses the integration circle. If the EDF used is such that PNR energies
diverge whenever a pole crosses the integration circle, it is important to note that
Variation After Projection (VAP) calculations will not converge as soon as the
minimization procedure "finds" the infinity~\cite{anguiano01b,stoitsov07a}.

All previous features prove that PNR calculations are ill-defined whenever poles
at $z\neq0$ arise and that the theory is unacceptable as it is. However, it is
possible to meaningfully regularize PNR calculations based on any EDF depending on
integer powers of the density matrices as was demonstrated in Paper~I and exemplified
in Paper~II. As a matter of fact, the method proposed in Paper~I precisely removes
the poles at $z=\pm z_{\mu}$ from $\mathcal{E}[z] \, \langle \Phi_1 | \Phi_{z} \rangle$.
However, it is crucial to realize that the correction method does not only remove those
poles but also consistently subtracts a spurious contribution to the physical pole
at $z=0$~\cite{bender07x}. In the end, only the physical pole at $z=0$ remains in
Eq.~(\ref{projenergy4}) and the independence of $\mathcal{E}^{N}$ on the integration
contour is recovered, as seen from Fig.~\ref{wellbehavedintegration2}; i.e.\ the same
PNR energy is obtained by integrating over circular contours $C_{R}$ of arbitrary
radius $R$.

\subsection{Non-integer power of densities}
\label{correctfractpower}

\subsubsection{Problem}
\label{problematic}

The situation is often more complex due to the presence of higher-order terms
of the form $\mathcal{E}^{\rho\rho\rho^{\alpha}}$ and
$\mathcal{E}^{\kappa\kappa\rho^{\gamma}}$ in realistic nuclear EDFs,
Eqs.~(\ref{fracrho}) and~(\ref{frackappa}).

If $\alpha = \gamma = 1$, then $\mathcal{E}^{\rho\rho\rho}$ and
$\mathcal{E}^{\kappa\kappa\rho}$ can, at least formally, be analyzed as if
they originated from a three-body vertex. Thus, and as for the bilinear terms,
two cases have to be distinguished (i) $\mathcal{E}^{\rho\rho\rho}$ and
$\mathcal{E}^{\kappa\kappa\rho}$ are both derived from the same antisymmetrized
three-body vertex and do not lead to divergences and steps in MR-EDF calculations
(ii) they refer to different three-body vertices such that the regularization
method proposed in Paper~I can be applied to obtain a meaningful PNR-EDF method.

However, all modern parameterizations of the nuclear EDF, starting either from
a functional approach or from a density-dependent vertex, depend on non-integer
powers of the density matrix that one cannot expand in a Taylor series to relate
them, at least formally, to three-body, four-body, \ldots forces. The goal
of the present paper is to characterize the pathologies brought about by such
dependencies and whether or not they are viable in the end; i.e.\ if the
corresponding pathologies can be easily regularized.

\subsubsection{Regularizing the integer part}
\label{regulintegerpart}

As a first step, one can reduce the extent of the problems associated with
terms of the form $\mathcal{E}^{\rho^{2m+n+\alpha}}$ and
$\mathcal{E}^{\kappa^{2m}\rho^{n+\gamma}}$, with $m$ and $n$ integer, and
$0 < \alpha < 1$ and $0 < \gamma < 1$, to pathologies only due to the
fractional powers $\rho^{\alpha}$ and $\rho^{\gamma}$, respectively. This means
that steps and potential divergences associated with the \emph{integer part}
$2m+n$ can be regularized from the outset. This is the case either (i) if one started
from a density-dependent $(2m+n)$-body effective force or (ii) by applying
the correction method proposed in Paper~I to $\mathcal{E}^{\rho^{2m+n}}$ and
$\mathcal{E}^{\kappa^{2m}\rho^{n}}$.

Let us exemplify how an empirical extension of the correction method proposed
in Paper~I can be designed to regularize the quadratic part of
$\mathcal{E}^{\rho\rho\rho^{\alpha}}$, with $0<\alpha<1$. To simplify the situation
further, we disregard the term $\mathcal{E}^{\kappa\kappa\rho^{\gamma}}$ in
the following discussion. Such a simplification does not alter any of the
conclusions given in the rest of the paper.

To proceed, we first introduce \emph{pseudo} two-body matrix elements
$\bar{v}^{\rho\rho\rho^{\alpha}}_{\mu\nu \mu\nu}[z]$ which take, for the
toy functional considered in the present paper, the form
\begin{widetext}
\begin{eqnarray}
\label{newmatrixelements}
\bar{v}^{\rho\rho\rho^{\alpha}}_{\mu\nu \mu\nu}[z]
& \equiv & 2  \int \! d^3r \;
     \big[ A^{\rho\rho\rho^{\alpha}} \, W^{\rho}_{\mu \mu} (\vec{r}) \
                                        W^{\rho}_{\nu \nu} (\vec{r})
         + A^{ss\rho^{\alpha}} \, \vec{W}^{s}_{\mu \mu} (\vec{r})
                            \cdot \vec{W}^{s}_{\nu \nu} (\vec{r})
     \big] \,
     \big[ \rho^{1z}(\vec{r})
     \big]^{\alpha}
\, .
\end{eqnarray}
\end{widetext}

With the pseudo two-body matrix elements $\bar{v}^{\rho\rho\rho^{\alpha}}_{\mu\nu
\mu\nu}[z]$ at hand, one can apply the correction formula given by Eq.~(43)
of Paper~II. However, and as opposed to terms of the EDF depending on integer
powers of the density matrices, the matrix elements of $\bar{v}^{\rho\rho\rho^{\alpha}}$
do depend on the gauge variable $z$. As a result, Eq.~(43) of Paper~II must be applied
in such a way that the matrix elements are located underneath the integral over $z$.
Last but not least, it would also be trivial to regularize the integer part of
$\mathcal{E}^{\kappa\kappa\rho^{\gamma}}$ by introducing the pseudo two-body matrix elements
$\bar{v}^{\kappa\kappa\rho^{\alpha}}[z]$ and by using them in Eq.~(43) of Paper~II.

\subsection{Left-over fractional power}
\label{problemfracpower}

With the latter correction at hand, the quadratic part of
$\mathcal{E}^{\rho\rho\rho^{\alpha}}$  does not create any divergence or step
in the PNR-EDF energy anymore. Again, the same is true if one starts from the
outset from a density-dependent two-body antisymmetrized \emph{interaction},
as long as the corresponding term $\mathcal{E}^{\kappa\kappa\rho^{\alpha}}$
is explicitly considered in the EDF  to proceed to the necessary recombination
of terms in Eq.~(\ref{enerpairproj1}). One way or another, one is only left in
the end with discussing the impact of the \emph{fractional} power of the
transition density; i.e.\ the extra factor $\big[\rho^{1z}(\vec{r})\big]^{\alpha}$,
with $0<\alpha<1$.

\subsubsection{Analytical structure of $\mathcal{E}^{\rho\rho\rho^{\alpha}}[z] \,
\langle \Phi_1 | \Phi_{z} \rangle$}

Now that the pathologies due to the bilinear factor in
$\mathcal{E}^{\rho\rho\rho^{\alpha}}$ have been taken care of,
the contribution of interest to the PNR energy can be written as
\begin{widetext}
\begin{equation}
\label{projfrac}
\mathcal{E}^{N}[\rho\rho\rho^{\alpha}]
\equiv \oint_{C_{R}} \frac{dz}{2i\pi c^{2}_{N}} \,
       \frac{\mathcal{E}^{\rho\rho\rho^{\alpha}} \left[z\right]}{z^{N+1}} \,
       \langle \Phi_1 | \Phi_{z} \rangle
\equiv \oint_{C_{R}} \frac{dz}{2i\pi c^{2}_{N}}
       \int \! d^3r \, \frac{F[z](\vec{r})}{z^{N+1}} \,
       \big[\rho^{1z}(\vec{r})\big]^{\alpha} \, ,
\end{equation}
where
\begin{eqnarray}
\label{functF}
F[z](\vec{r})
& \equiv& z^{4} \sum_{\nu\neq\mu,\bar{\mu}}
          \big[  A^{\rho\rho\rho^{\alpha}} \,
                 W^{\rho}_{\mu \mu} (\vec{r}) \, W^{\rho}_{\nu \nu} (\vec{r})
               + A^{ss\rho^{\alpha}} \,
                 \vec{W}^{s}_{\mu \mu} (\vec{r}) \cdot \vec{W}^{s}_{\nu \nu} (\vec{r})
          \big] \, v^{2}_{\mu} \, v^{2}_{\nu}
          \prod_{\zeta >0   \atop \zeta \neq \mu, \nu}
          (u_{\zeta}^2 + v_{\zeta}^2 \, z^{2}) \, ,
\end{eqnarray}
\end{widetext}
with $N$ even. In agreement with the properties of $\mathcal{E}[z]$
mentioned above, $F[z](\vec{r})$ is an even function of $z$ for all
$\vec{r}$. For odd $N$, it is easy to prove that $F[z](\vec{r})$ is
an odd function of $z$ in such a way that $F[z](\vec{r})/z^{N+1}$
remains itself an odd function of $z$.

The terms corresponding to $\nu=\mu$ and $\nu=\bar{\mu}$ are absent in
Eq.~(\ref{functF}) because (i) they were removed by the correction method
briefly outlined in Sect.~\ref{correctfractpower} (ii) one started from a
density-dependent two-body interaction; i.e.\ the term with $\nu=\mu$ do
disappear ($A^{ss\rho^{\alpha}}=-A^{\rho\rho\rho^{\alpha}}$) whereas the
term with $\nu=\bar{\mu}$ could be combined with the corresponding one
in $\mathcal{E}^{\kappa\kappa\rho^{\alpha}}$ to give a well-behaved
contribution that we omit here.

\begin{figure}[t]
\includegraphics[height=10.cm,angle=-90]{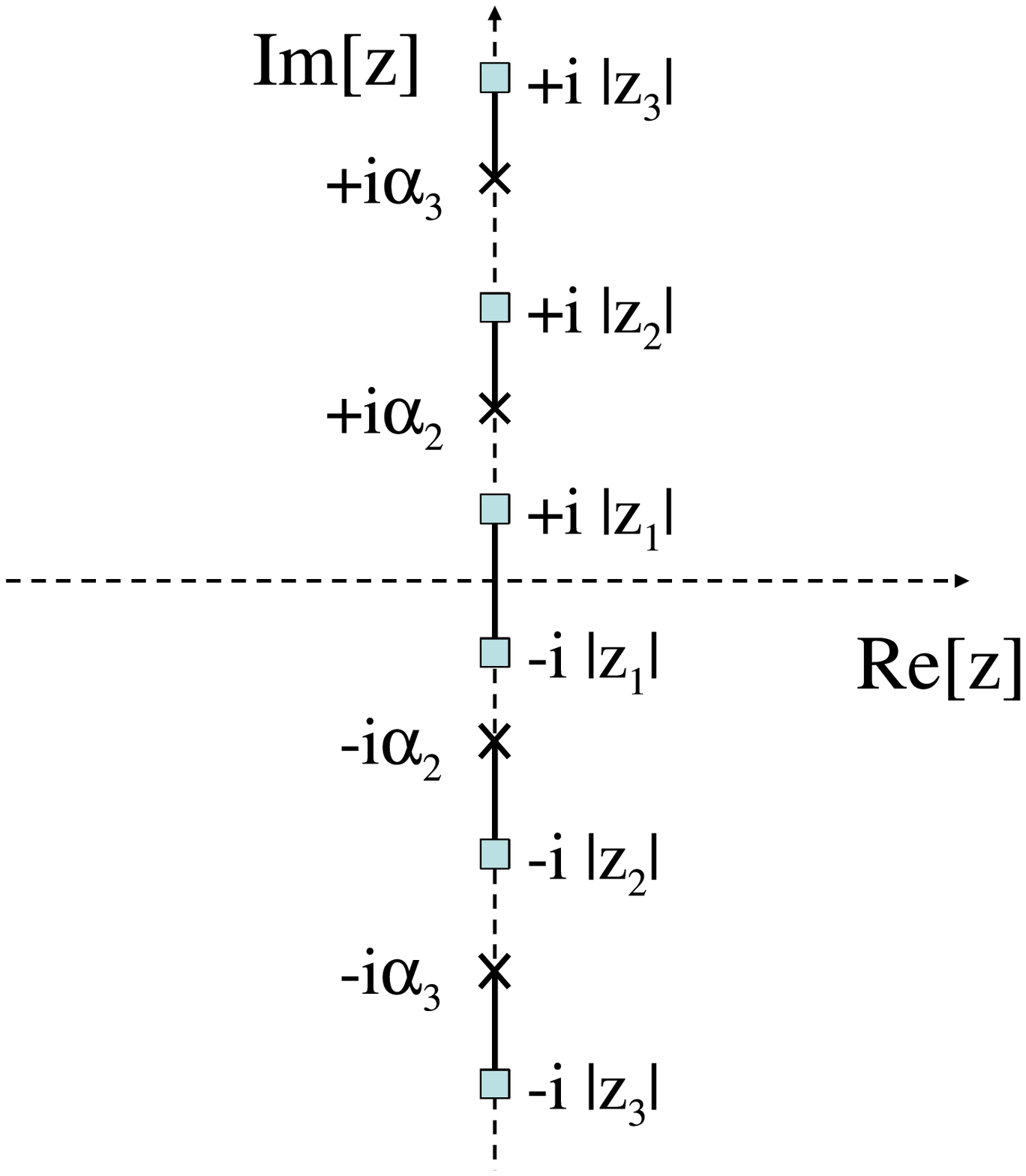}
\caption{\label{branchcuts}
Branch cuts of $\big[\rho^{1z}(\vec{r})\big]^{\alpha}$. The branch cuts
join the integrable poles of $\big[\rho^{1z}(\vec{r})\big]^{\alpha}$ at
$z=\pm i |u_{\mu}/v_{\mu}|$ (squares) and its zeros at
$z=\pm i \alpha_{\mu}$ (crosses).
}
\end{figure}

To understand the features displayed by the contribution
$\mathcal{E}^{N}[\rho\rho\rho^{\alpha}]$ to the PNR energy, it is necessary
to extract for each $\vec{r}$ the non-analytical structure of the integrand
in Eq.~(\ref{projfrac}) where the order of the two integrals over $\vec{r}$
and $z$ have been reversed. Clearly, the function $F[z](\vec{r})/z^{N+1}$
displays a (physical) pole at $z=0$. The difficulty comes from the fractional
power of the local transition density that multiplies $F[z](\vec{r})$. Indeed,
such a function is multivalued on the complex plane for all $\vec{r}$.

Defining the function corresponding to taking the fractional power of a complex
number\footnote{Parameterizing $z=re^{i\theta}$, $\theta \in [-\pi, + \pi]$,
we define the principal value of the function $z^{\alpha}$, $\alpha$ being a
rational number between zero and one, as $z^{\alpha}\equiv r^{\alpha}
e^{i\alpha\theta}$. The latter choice lifts the ambiguity regarding the
multivalued nature of the function but requires to track the latter through
several Riemann cuts.} requires the introduction of a branch cut along the
axis where that number is real and negative. Here, this means that one needs
the values of $z$ for which the function $\rho^{1z}(\vec{r})$ is real and
negative. As can be seen from Eqs.~(\ref{eq:locrho}) and~(\ref{contractphcomplex}),
the transition density is real both on the real and imaginary axis, but can be
negative only on the latter. A discussed in Ref.~\cite{doba05a}, $\rho^{1z}(\vec{r})$
is negative for $z=iy$ such that $|z_{\mu-1}| < \alpha_{\mu} < y < |z_{\mu}|$,
as well as on the entire interval $[-z_{1},+z_{1}]$, where $z_{1}$ denotes
the closest pole to the origin. The corresponding branch cuts are characterized
in Fig.~\ref{branchcuts} by solid lines joining the zeros of $\rho^{1z}(\vec{r})$
at $z=\pm i \alpha_{\mu}$ (crosses) and its next integrable pole at
$z=\pm z_{\mu}$ (square). Whereas the poles of $\rho^{1z}(\vec{r})$ are
independent of the position vector $\vec{r}$, the points $z=\pm i \alpha_{\mu}$
at which it changes sign in between two poles do depend on~$\vec{r}$.

\subsubsection{Calculation of $\mathcal{E}^{N}[\rho\rho\rho^{\alpha}]$}
\label{integraleplancomplex}

\begin{figure}[t]
\includegraphics[height=10.cm,angle=-90]{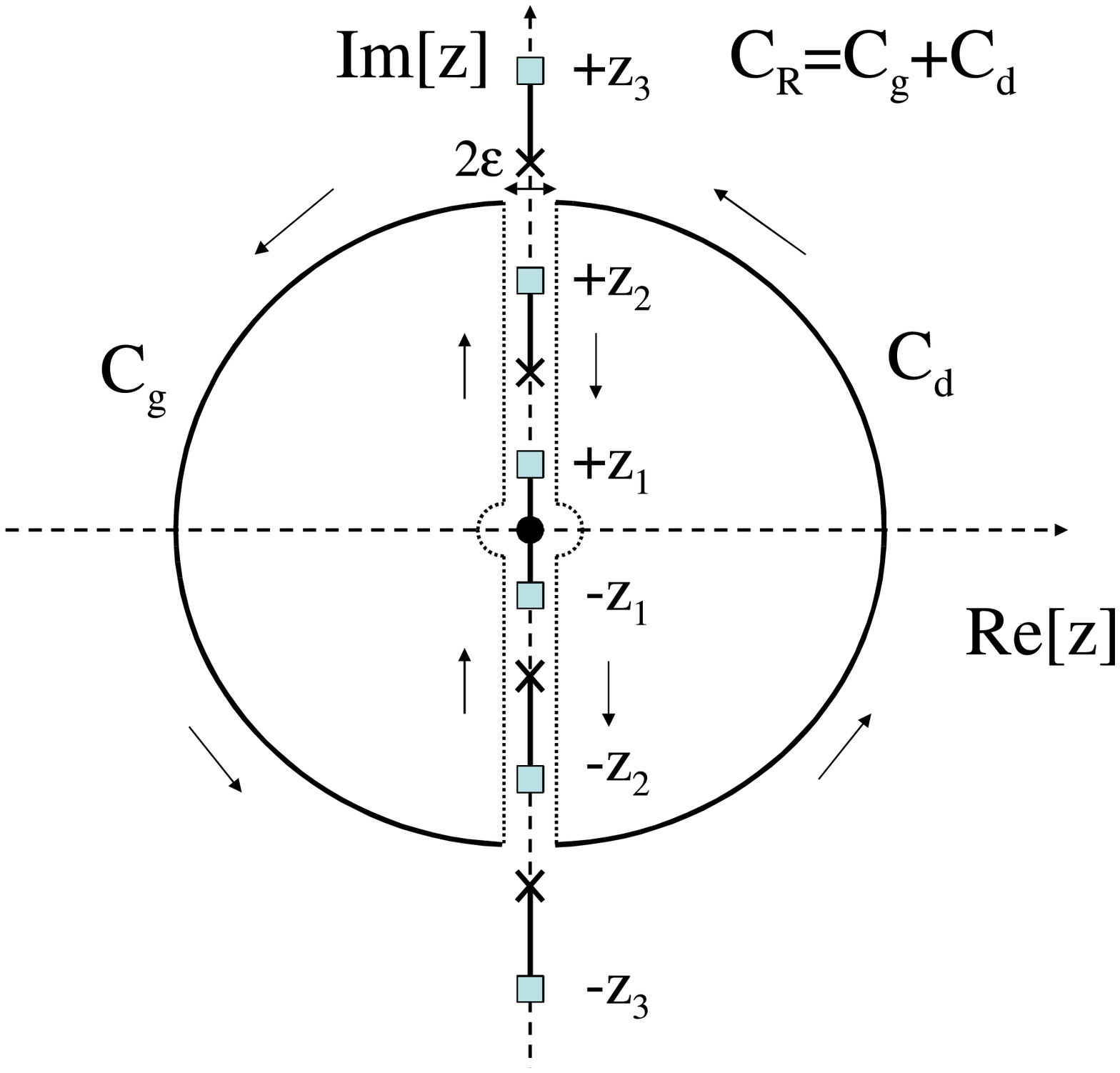}
\caption{\label{integfrac}
Specification of the integration contour for an EDF containing
fractional powers of the densities.
}
\end{figure}

Knowing the non-analytical structure of the integrand
$F[z](\vec{r})\big[\rho^{1z}(\vec{r})\big]^{\alpha} / z^{N+1}$, the
integration contour to be used in Eq.~(\ref{projfrac}) can be specified.
Just as before, the circle $C_{R}$ needs to be deformed in order to apply
the Cauchy theorem on contours encircling regions where the function is
entirely analytical. In particular, one cannot go through branch cuts as
one must remain on the same Riemann sheet. An acceptable decomposition
under the form $C_{R}\equiv \big[C_{g}+C_{d}\big](\epsilon \rightarrow 0)$,
where each semi-circle $C_{g}/C_{d}$ is further closed by a vertical
segment along the imaginary axis interrupted by a semi-circle around the
origin, is displayed in Fig.~\ref{integfrac}. Note that, as opposed to
Fig.~\ref{wellbehavedintegration}, no special care needs to be taken
around the poles at $z=\pm z_{\mu}$ as they are now integrable
($\sim 1/z^{\alpha}$ with $0< \alpha<1$). The crucial point, however,
is that the portions along the branch cuts will not cancel out as we sum
the two vertical segments because the integrand (in fact
$\big[\rho^{1z}(\vec{r})\big]^{\alpha}$) is discontinuous across the
branch cuts.

\begin{figure}[t]
\includegraphics[height=9.cm,angle=-90]{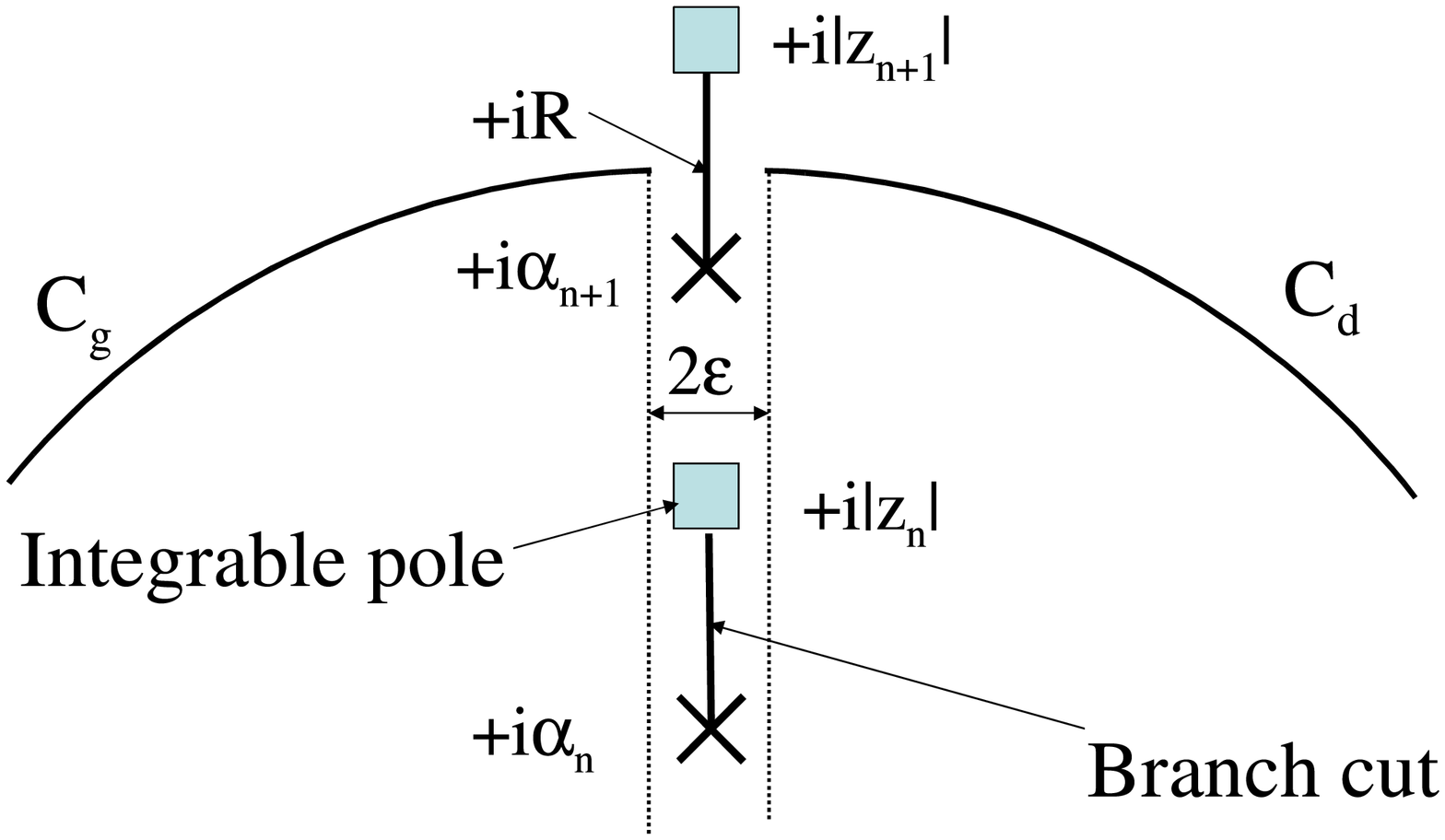}
\caption{\label{integfraczoom} Zoom on the integration contour $\mathcal{C}_{R}$
obtained as the limit of the sum of two disconnected semi-circles. For illustration,
we display a situation where the chosen integration contour $C_{R}$ "hits" the
$(n+1)^{th}$ branch cut at $z=\pm i R$, that is, has a radius $R$ such that
$\alpha_{n+1}\leq  R \leq |z_{n+1}|$.}
\end{figure}

One may wonder what happens when, as in Fig~\ref{integfraczoom}, the
radius $R$ is such that the original contour $C_{R}$ goes through a branch
cut. In fact, the contour $C_{R}$ defined through $\big[C_{g}+C_{d}\big]
(\epsilon \rightarrow 0)$ in Fig.~\ref{integfrac} (i) is well defined when
a branch cut lies in between $C_{g}$ and $C_{d}$ because the limit
$\epsilon \rightarrow 0$ does not pose any problem once the value of
the function on both sides of the cut has been properly worked out, (ii)
is the contour which has been used in actual
calculations~\cite{bender03c,anguiano01b,doba05a} and (iii) might however
need to be discretized on a rather dense mesh to provide converged calculations.

Note that the deformation of the contour discussed above was advocated
in Ref.~\cite{doba05a} as a remedy to the pathology brought about by
branch cuts. In fact, it is rather a necessary step to simply \emph{define}
the integration over the original circle and obtain the result it provides.
As detailed below, proceeding to such a deformation of the contour does not
remove the intrinsic pathological nature of MR calculations performed using
an EDF containing non-integer powers of the density matrices.

We are now ready to apply the Cauchy theorem along the two closed contours
appearing in Fig.~\ref{integfrac} and then let $\varepsilon$ goes to zero. It
is clear that the contributions from the vertical portions in between the
branch cuts cancel out as we add the results from the two closed contours.
On the other hand, contributions from segments along the branch cuts will not
cancel out because of the discontinuity of the integrand across them.

\begin{figure}[t]
\includegraphics[height=10.cm,angle=-90]{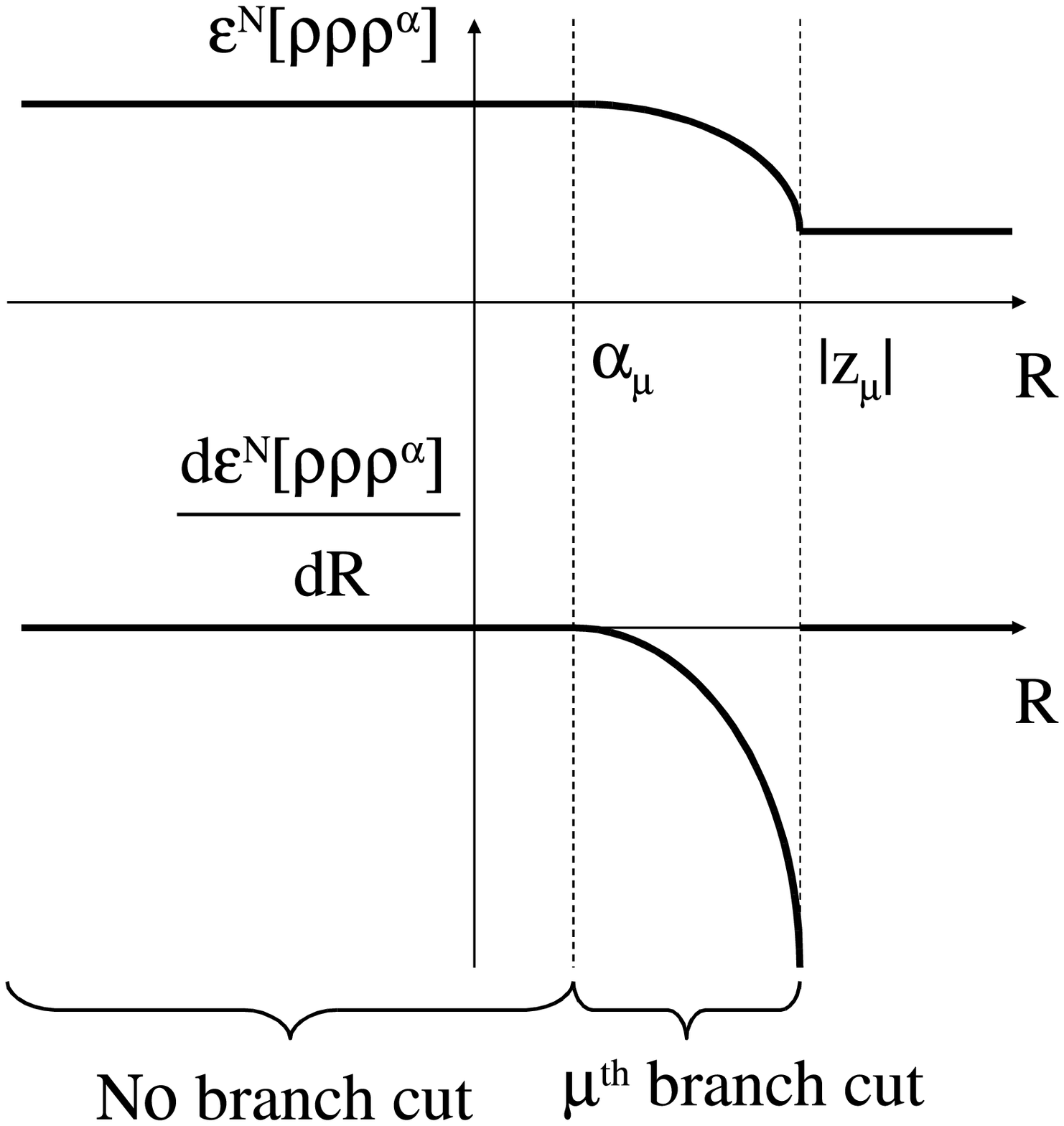}
\caption{\label{pointanguleux}
Schematic effect of a shift transformation on the PNR energy. Top: projected
energy $\mathcal{E}^{N}[\rho\rho\rho^{\alpha}]$ as a function of $R$. Bottom: same
for the derivative of  $\mathcal{E}^{N}[\rho\rho\rho^{\alpha}]$ with respect to $R$.
}
\end{figure}

We consider for illustration (see Fig.~\ref{integfraczoom}) the situation where
the contour $C_{R}$ "hits" the $(n+1)^{th}$ branch cut at $z=\pm iR$; i.e.\
$\alpha_{n+1}\leq R \leq |z_{n+1}|$. This means that the $n^{th}$ branch cut
is entirely located inside $C_{R}$ whereas the $(n+1)^{th}$ one is partially
outside the circle of integration. For simplicity, and because it is irrelevant
to the present discussion, we do not calculate the contribution
$\mathcal{E}^{N}[\rho\rho\rho^{\alpha}]([-z_{1}, + z_{1}])$ from the
closest branch cut to the origin. Indeed, this one is trickier than the
other branch cuts because the physical pole at $z=0$ lies on that branch
cut. All that matters for the present discussion is that the branch cut
$[-z_{1}, + z_{1}]$ provides a finite contribution to the projected energy.
In the end, one obtains
\begin{widetext}
\begin{eqnarray}
\label{projfrac2}
\mathcal{E}^{N}[\rho\rho\rho^{\alpha}](R) - \mathcal{E}^{N}
[\rho\rho\rho^{\alpha}]([-z_{1}, + z_{1}])
& = & (-1)^{\frac{N}{2}} \, \frac{2}{\pi} \, \sin(\alpha\pi)
      \left[\sum_{\mu=1}^{n}  \int_{\alpha_{\mu}}^{|z_{\mu}|}\!\! dy
                           + \int_{\alpha_{n+1}}^{R}\!\! dy \,
      \right]
      \int \! d^3 r \, \frac{F[iy](\vec{r})}{y^{N+1}} \,
      \Big|\rho^{1 \, iy}(\vec{r})\Big|^{\alpha}
      \, ,
\end{eqnarray}
which is real and where, for $y$ real,
\begin{eqnarray}
\rho^{1 \, iy}(\vec{r})
& = & \sum_{\mu} W^{\rho}_{\mu\mu}(\vec{r}) \, \frac{y^{2}}{ y^{2} - |z_{\mu}|^{2} }  \, ,
      \\
\label{functF2}
F[iy](\vec{r})
& = & y^{4} \sum_{\nu\neq\mu,\bar{\mu}} \Big[A^{\rho\rho\rho^{\alpha}} \, W^{\rho}_{\mu \mu} (\vec{r}) \,
      W^{\rho}_{\nu \nu} (\vec{r}) + A^{ss\rho^{\alpha}} \,  \vec{W}^{s}_{\mu \mu} (\vec{r}) \cdot
     \vec{W}^{s}_{\nu \nu} (\vec{r}) \Big] \, v^{2}_{\mu} \, v^{2}_{\nu} \,
     \prod_{\zeta >0   \atop \zeta \neq \mu, \nu} (u_{\zeta}^2 - v_{\zeta}^2 \, y^{2}) \, .
\end{eqnarray}
\end{widetext}
The above analytical results are explicit enough that we can draw several important
conclusions from them. First, Eq.~(\ref{projfrac2}) demonstrates that the PNR energy
depends on the radius $R$ of the integration contour through the boundary of the
integral; i.e.\ the PNR energy is not shift invariant. As $C_{R}$ goes through a
branch cut, the contribution of that branch cut changes progressively and leaves
a smoothed step in the PNR energy; see Fig.~\ref{pointanguleux}. This relates to an
unphysical breaking of shift invariance. Second, there is no discontinuity or
divergence as $C_{R}$ passes through the branch points since the function
$|\rho^{1 \, iy}(\vec{r})|^{\alpha}$ is integrable at $y=|z_{\mu}|$, for all $\mu$.

The two previous conclusions are at variance with what happens for (most of the)
EDFs containing only integer powers of the densities as recalled in
Sect.~\ref{correctbilinear}. Indeed, a pole crossing the integration provides in
this case PNR energies with (i) an abrupt step (ii) a divergence if the pole is of
even order~\cite{bender07x}. Also, it is important to underline the role played
by the regularization of the bilinear part of $\mathcal{E}^{\rho\rho\rho^{\alpha}}$
put forward in Sect.~\ref{regulintegerpart}. If one were to use the
\emph{uncorrected} term $\mathcal{E}^{\rho\rho\rho^{\alpha}}$, the PNR energy
would diverge as $C_{R}$ passes through the branch points. Indeed, the integrand
in Eq.~(\ref{projfrac2}) would then contain terms overall proportional to
$(y^{2} - |z_{\mu}|^{2})^{-1}|\rho^{1 \, iy}(\vec{r})|^{\alpha}$ which is
\emph{not} integrable at $y=|z_{\mu}|$.

In any case, the absence of divergence for the \emph{regularized}
$\mathcal{E}^{N}[\rho\rho\rho^{\alpha}]$ is critical since the associated
integrability of the pole was used in Ref.~\cite{anguiano01b} to assess the
meaningfulness of PNR calculations performed with the Gogny force. However,
and although divergences do constitute a dramatic pathology of ill-defined PNR
calculations, the most profound problem relates rather to the breaking of
shift invariance of the PNR energy as one changes the integration contour.
Indeed, the associated spurious branch cuts modify the topology of potential
energy curves as one deforms the system with respect to a collective degree of
freedom. As discussed above, such a problem persists for a regularized
non-integer power or, equivalently, for an effective two-body vertex depending
on a fractional power of the density. Still, the absence of divergence explains
why the spurious nature of fractional powers of the densities that we focus on
here has been overlooked so far even more than the pathologies brought about by
integer powers.

In the end, divergences are presently replaced by another pathological behavior
of the PNR energy. To isolate such a pattern, let us take the derivative of
$\mathcal{E}^{N}[\rho\rho\rho^{\alpha}](R)$ in Eq.~(\ref{projfrac2}) with
respect to the radius of integration $R$. One obtains, for $\mu > 1$
\begin{widetext}
\begin{equation}
\label{derivee2}
\frac{d \mathcal{E}^{N}[\rho\rho\rho^{\alpha}]}{dR}\Bigg|_{R}
= \left\{
  \begin{array}{ll}
     0 & \text{if $R \in \big[|z_{\mu-1}|, \alpha_{\mu}\big]$ ,}  \\
    {\displaystyle
    \frac{(-1)^{N/2}}{R^{N+1}} \, \frac{2}{\pi} \, \sin(\alpha\pi) \,
    \int \! d^3r \; F[iR](\vec{r}) \,
    \big|\rho^{1 \, iR}(\vec{r})\big|^{\alpha} } \qquad
    & \text{if $R \in \big[\alpha_{\mu},|z_{\mu}|\big]$.}
  \end{array}
\right.
\end{equation}
\end{widetext}
Because of the non-analytic behavior of $\big|\rho^{1 \, iR}(\vec{r})\big|^{\alpha}$
at each branch point, the derivative diverges in Eq.~(\ref{derivee2}) for
$R=|z_{\mu}|$, $\mu \neq 1$. As a result, the projected energy displays a
\emph{kink} (non-derivable behavior) as the integration circle goes through a
branch point or as a branch point goes through the integration circle when the
system is deformed along a collective path. This fact alone is unacceptable for
a well-defined projected theory. The corresponding pattern is schematically
displayed in Fig.~\ref{pointanguleux} and is observed in realistic calculations
as  will be discussed in Sect.~\ref{applications}.

\subsection{Isospin degree of freedom}
\label{isospin}

The isospin degree of freedom does not modify any conclusion of the present
paper but only complexifies certain aspects of the discussion. Still, to
provide an idea of the modifications brought about by the consideration of
both protons and neutrons, we now proceed to a restricted set of remarks.

Considering the isospin degree of freedom, one must account for the fact
that densities, e.g.\ $\rho_q (\vec{r})$, and single-particle wave-functions $\varphi_{\mu} (\vec{r}q)$ are now labeled with the isospin projection quantum number $q$, where $q=n$ and $q=p$ for neutrons and protons, respectively. The problematic terms entering the toy Skyrme functional (Eqs.~\ref{fracrho}-\ref{frackappa}) now take the form
\begin{eqnarray}
\label{fracrho2}
\mathcal{E}^{\rho\rho\rho^{\alpha}}
& \equiv & \int \! d^3r \! \sum_{q = p,n}
           \Big[  A^{\rho\rho\rho^{\alpha}} \rho_q^2 (\vec{r})
                + A^{ss\rho^{\alpha}}       \vec{s}_q^2 (\vec{r})
           \Big] \, \rho^{\alpha}_{0}(\vec{r})
           \nonumber \\
&        & + \int \! d^3 r \! \! \! \sum_{q , q' = p,n \atop q \neq q'} \! \!
           \Big[  B^{\rho\rho\rho^{\alpha}} \rho_q (\vec{r}) \, \rho_{q'} (\vec{r})
           \\
&        & \phantom{\int \! d^3 r \! \! \! \sum_{q , q' = p,n \atop q \neq q'}  \! \! \Big[ }
                + B^{ss\rho^{\alpha}}       \vec{s}_q (\vec{r}) \cdot \vec{s}_{q'} (\vec{r})
           \Big] \, \rho^{\alpha}_{0}(\vec{r}) \, ,
           \\
\label{frackappa2}
\mathcal{E}^{\kappa\kappa\rho^{\gamma}}
& \equiv & \int \! d^3r \! \sum_{q = p,n}  A^{\tilde{\rho} \tilde{\rho}\rho^{\gamma}} \,
           \left|\tilde{\rho}_{q} (\vec{r})\right|^{2} \rho^{\gamma}_{0}(\vec{r}) \, ,
\end{eqnarray}
where the coupling constants $A/B$ characterize terms in which the two linear
densities involved refer to identical/different isospins. Note that neutron-proton
pairing is not considered. Also, $\rho_{0}(\vec{r})$ is the isoscalar part of the
matter density. As single-particle states have a definite isospin projection,
$\rho_{0}(\vec{r}) = \rho_{n}(\vec{r})+\rho_{p}(\vec{r})$.

In the present case, both neutron and proton particle numbers are restored. Doing
so requires to consider two gauge angles $\varphi_{n}$ and $\varphi_{p}$ for
neutrons and protons, respectively. As a result, PNR energies are obtained
through a double integration over the complex plane where the corresponding
variables are denoted as $z_{n}$ and $z_{p}$.

As far as the regularization of the bilinear part of the toy functional, see
Sect.~\ref{regulintegerpart}, it still leads to the condition $A^{ss} = - A^{\rho\rho}$
and thus only constrains the like-particle interaction. Then, one notes that the
pseudo matrix elements introduced in Eq.~(\ref{newmatrixelements}) to deal with the
part of the EDF containing non-integer powers of the density matrices now depend
on both the neutron $z_{n}$ and proton $z_{p}$ gauge variables because of the
dependence on the isoscalar part of the transition local density in
Eqs.~(\ref{fracrho2}-\ref{frackappa2}). With the pseudo two-body matrix elements
$\bar{v}^{\rho\rho\rho^{\alpha}}_{\mu\nu \mu\nu}[z_{n},z_{p}]$ at hand, one can
apply the correction formula of Eq.~(43) of Paper~II ensuring that the matrix
elements are now placed underneath the integrals over the two gauge angles.

Once the part of the energy kernel $\mathcal{E}[z_{n},z_{p}]$ that depends
only on integer powers of the density matrix has been regularized, one is
left with the spuriosities brought by the fractional power of the isoscalar
transition density $\big[\rho^{1z_{q}}_{q}(\vec{r})
+\rho^{1z_{\bar{q}}}_{\bar{q}}(\vec{r})\big]^{\alpha}$. The branch cuts of
the latter are not the same as those seen when dealing with a single particle
species. This modifies the analysis but does not change the fact that the
theory is not satisfactory, irrespective of the fine tuning done to define
the integration contour. As a result, PNR energies cannot be made shift
invariant and display smooth spurious steps as one changes the proton
and/or neutron radii of integration or deforms the system along a certain
degree of freedom.

\section{Applications}
\label{applications}

We wish to illustrate the analytical results obtained in the previous Sections
through results of realistic calculations. We perform PNR calculations after
variation of \nuc{18}{O}. We use the SLy4 parametrization~\cite{chabanat98} of
the Skyrme EDF together with a pairing functional derived from a Delta
Interaction (DI). The Coulomb exchange part of the functional, usually calculated
in the Slater approximation, is omitted as done in Paper~II. The SLy4
Skyrme parametrization includes a term of the type $\mathcal{E}^{\rho\rho\rho^{1/6}}$
which is perfectly suited to the present discussion.

\subsection{Uncorrected calculations}
\label{uncorrected}

\begin{figure}[t]
\includegraphics[width=8.0cm]{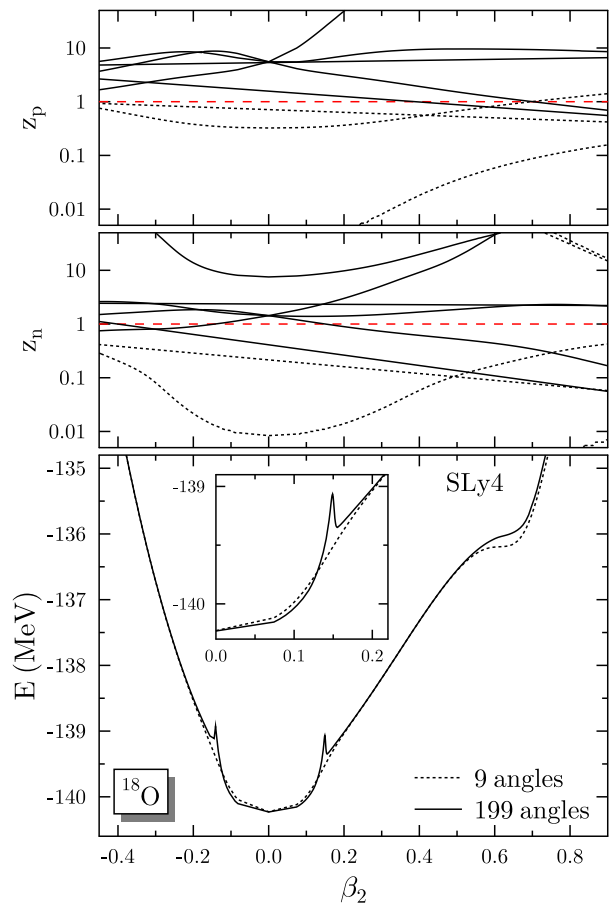}
\caption{\label{uncorrectedPES} (Color online) Spectrum of poles $z_\mu = |u_\mu/v_\mu|$ for protons (top panel) and neutrons (middle panel) as a function of quadrupole
deformation, which for levels in the vicinity of the Fermi energy resembles a stretched and slightly distorted Nilsson diagram. The dashed red line at $z_q = 1$ denotes the radius of the standard integration-contour $R_q = 1$. The bottom panel shows the PNR energy $\mathcal{E}^{N}$ for two different numbers of discretization points in the
computation of the integrals over the gauge neutron $\varphi_{n}$ and proton
$\varphi_{p}$ angles.}
\end{figure}

As explained in Sect.~\ref{notations}, traditional PNR calculations have been
performed using non-diagonal kernels defined through the prescription
$\mathcal{E}[0,\varphi] \equiv \mathcal{E}[\rho^{0\varphi},\kappa^{0\varphi},
\kappa^{\varphi 0 \, \ast}]$, where $\mathcal{E}[\rho,\kappa,\kappa^{\ast}]$
is the single-reference EDF. Figure~\ref{uncorrectedPES} shows the PNR energy
$\mathcal{E}^{N}$ obtained in this way for \nuc{18}{O} and displayed as a
function of quadrupole deformation. The calculation is repeated twice, using
5 and 199 points in the discretization of the integrals over the two gauge angles.

One observes that the deformation energy surface obtained with 5 integration points
is smooth and looks physically reasonable. However, as one increases the number of
integration points, divergences develop, precisely at deformations where a neutron
or a proton single-particle state crosses the Fermi energy in the underlying SR
states, i.e.\ when the associated non-integrable branch point crosses the unit circle
in the complex plane. This is consistent with the discussion given in
Sect.~\ref{integraleplancomplex} for the \emph{uncorrected} SLy4 parametrization.
Such divergences are at variance with the results obtained in Paper~II with the
SIII parametrization. Indeed, SIII is of specific functional form such that all
the poles at $z=\pm z_{\mu}$ are simple poles. This is notably due to the fact that
the trilinear terms entering SIII do not display products of three density matrices
referring to the same isospin. As explain in Sect.~\ref{projEintegerpowers}, this
property leads to a finite Cauchy principle value as the poles cross the integration
circle.

\begin{figure}[t]
\includegraphics[width=8.0cm]{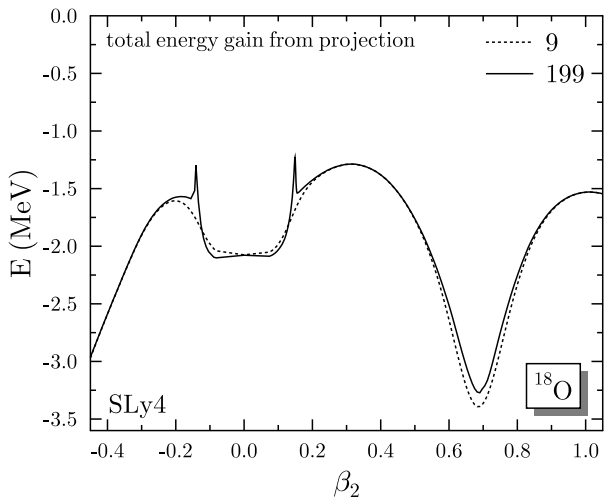}
\caption{\label{correctedPES1} Energy gain from particle number restoration
as a function of quadrupole deformation for two different numbers of discretization
points in the computation of the integrals over the gauge angles.
}
\end{figure}

Still, the finite step left in the PNR energy as a pole/branch cut enters or leaves the
integration circle is a pathology shared by the calculations performed with SLy4 and SIII.
Those steps are better visible in Fig.~\ref{correctedPES1} which displays the gain from
particle number restoration with respect to the SR energy (rather than the absolute PNR
binding energy) using SLy4. Note in passing that the reason why the structure around
$\beta_{2}=0.7$ does not display a typical step can be understood from the fact that
two pairs of levels cross the Fermi energy at that deformation, as discussed in Paper~II.

By looking carefully, one can observe an interesting difference between the steps
produced by SIII (see Paper~II) and those obtained presently using SLy4. The steps
generated by SLy4 are significantly less steep than those produced by SIII. This is
because, whereas a sharp step is generated by an isolated pole leaving or entering
the integration circle in the case of SIII, which occurs over an infinitesimal interval
of deformation, it is generated by a branch cut leaving or entering the integration
circle in the case of SLy4, which happens over a finite interval of deformation.

\subsection{Correcting the bilinear part} \label{quadraticcorrected}

The specificity of SLy4 is to contain a term of the type $\mathcal{E}^{\rho\rho\rho^{1/6}}$.
As discussed in Sect.~\ref{correctfractpower}, one could have hoped that regularizing the
quadratic part of this term through the correction method proposed in Paper~I would lead
to a well-behaved PNR energy; i.e.\ that the remaining fractional power of the density
would not create any pathology, in particular in view of the fact that the branch point
becomes integrable in this case. Of course, it is important to remember that the correction
method proposed in Paper~I relies on solid basis only for terms of the form
$\mathcal{E}^{\rho^{n}}$, with $n$ integer. Thus, regularizing the quadratic part of
$\mathcal{E}^{\rho\rho\rho^{1/6}}$ in this way is purely empirical.

\begin{figure}
\includegraphics[width=8.0cm]{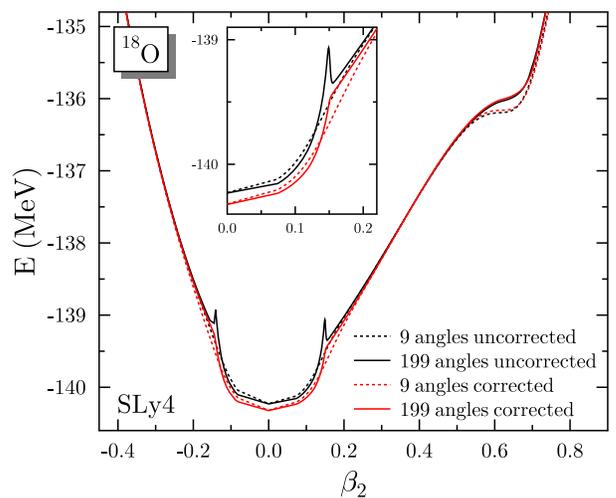}
\caption{\label{correctedPES2}(Color online)  Particle number restored energy $\mathcal{E}^{N}$ as
a function of quadrupole deformation without and with regularization of
all bilinear terms in the EDF, including the quadratic part of
$\mathcal{E}^{\rho\rho\rho^{1/6}}$. Results are shown for two different
numbers of discretization points in the computation of the integrals
over the gauge angles.}
\end{figure}

\begin{figure}[t]
\includegraphics[width=8.0cm]{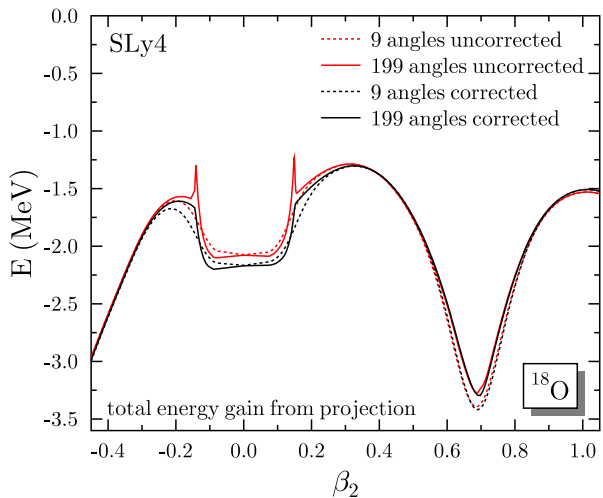}
\caption{\label{correctedPES3}(Color online)  Energy gain from PNR as a function of quadrupole
deformation without and with regularization of all bilinear terms in
the EDF, including the quadratic part of $\mathcal{E}^{\rho\rho\rho^{1/6}}$.
Results are shown for two different numbers of discretization points in
the computation of the integrals over the gauge angles.
}
\end{figure}

As a matter of fact, the results displayed in Fig.~\ref{correctedPES2} demonstrate
that proceeding to such a correction does not lead to a well-behaved PNR energy. The
integrability of the branch points remaining after regularizing the quadratic part of
$\mathcal{E}^{\rho\rho\rho^{1/6}}$ is such that all the divergences have disappeared.
This is a necessary but not sufficient condition to obtain a well-behaved PNR energy.
Indeed, Fig.~\ref{correctedPES3} clearly demonstrates that the spurious steps are still
present and have in fact not been reduced by regularizing the bilinear part of
$\mathcal{E}^{\rho\rho\rho^{1/6}}$. In addition, one observes that the corrected results
still depend strongly on the discretization of the integrals over the gauge angles. More precisely, all terms of the energy functional that are strictly bilinear have become independent on the number of discretization points whereas the term with the extra fractional power is not.
Considering the experience we have gathered about well-behaved PNR energies, such a
dependence is a fingerprint of a ill-defined PNR theory.

As discussed in Sect.~\ref{integraleplancomplex}, Figs.~\ref{correctedPES2}
and~\ref{correctedPES3} also show that regularizing the quadratic part of
$\mathcal{E}^{\rho\rho\rho^{1/6}}$ leads to the replacement of divergences by
non-derivable points in the PNR potential energy curve. Indeed, kinks are
clearly visible at the deformation where the divergences appeared before
applying the correction method. Using more mesh points for $Q_{20}$,
$\varphi_{p}$ and $\varphi_{n}$, one could resolve even better the non-derivable
character of the energy as a branch point passes through the integration circle.
This pattern relates directly to the analytical result obtained in Eq.~(\ref{derivee2}).

Finally, note that it is a particularity of the SLy4 interaction complemented with the
pairing interaction chosen here that the combined correction of all
density-independent terms is always very small in \nuc{18}{O}, often
even difficult to resolve on the plots.

\subsection{Shift transformation} \label{shifttransformation}

The finite steps that arise in the deformation energy surface are a reminiscence
of the violation of the shift invariance of the PNR energy. Such a violation is
unambiguously demonstrated by varying the radius of the integration contour
in Eq.~(\ref{projfrac2}); i.e.\ by computing Eq.~(\ref{derivee2}) as a function of $R$.

\begin{figure}
\includegraphics[width=8.0cm]{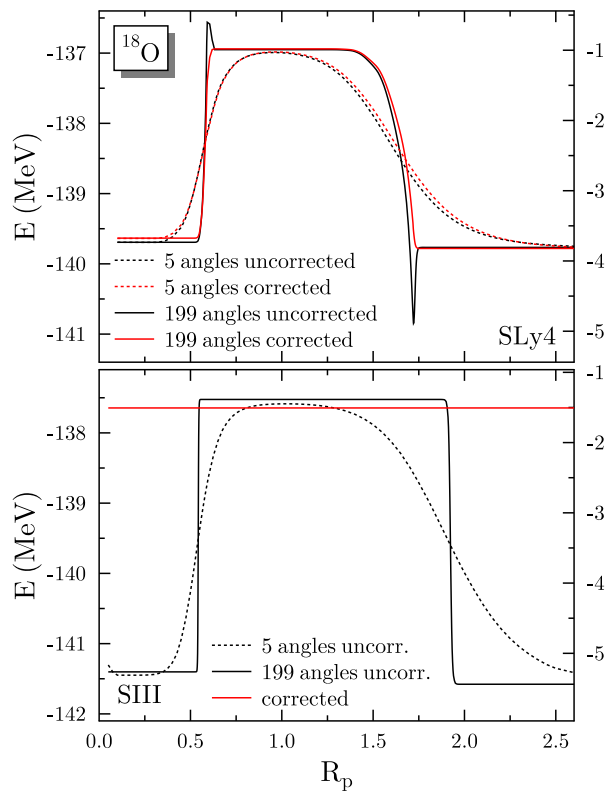}
\caption{\label{shift}
(Color online) Particle-number restored energy $\mathcal{E}^{N}$ as a function of the radius $R_{p}$
of the contour chosen to restore proton number ($R_{n}=1$) and for two different
numbers of discretization points in the computation of the integrals over the gauge
angles. Results are shown before and after regularization of the bilinear part of the
EDF. Upper panel: at a prolate quadrupole deformation $Q_{20} = 600$ fm$^2$ using the SLy4
parametrization. Bottom panel: at a prolate quadrupole deformation $Q_{20} = 500$ fm$^2$
using the SIII parametrization. The corrected SIII curve is independent on the number of discretization point; hence, only one
curve is shown. The left scale shows the absolute value of the
binding energy whereas the right scale shows the energy gain from symmetry restoration.
}
\end{figure}

The upper panel of Fig.~\ref{shift} shows the PNR energy of \nuc{18}{O} at a
deformation $Q_{20} = 600$ fm$^2$, obtained using the SLy4 parametrization. The
energy is displayed as a function of the radius of the integration contour used
to restore the proton number. The radius for the neutrons is $R_n = 1$ in all
cases. The calculation is performed with and without a regularization of the bilinear
part of the functional and for two different numbers of integration points (taken to
be the same for protons and neutrons). Finally, the bottom panel of Fig.~\ref{shift}
shows the same quantity obtained from the SIII parametrization at a quadrupole
deformation $Q_{20} = 500$ fm$^2$.

The upper panel of Fig.~\ref{shift} confirms that, even after regularizing the bilinear
part of $\mathcal{E}^{\rho\rho\rho^{1/6}}$, the PNR energy is not invariant under shift
transformation. Even though the correction method does remove the divergence, it does
not eliminate the shaped steps as the integration contour goes through a branch cut. In
addition, both the corrected and uncorrected PNR energies depends strongly on the
discretization of the integrals. Again, those two features are entirely due to the term in the functional depending on a non-integer power of the density. After regularization, all terms that are strictly bilinear become shift invariant. For comparison, the bottom panel of Fig.~\ref{shift}
shows the PNR energy obtained with SIII in Paper~II. We recall that SIII contains only
linear, bilinear and trilinear terms which are such that all poles at $z=\pm z_{\mu}$
are of order one. The corresponding PNR energy is, after regularization, independent on
the contour and the number of discretization points with a numerical precision better
than 1~keV. When
restoring the particle number that the SR-EDF calculation was constrained to, the finite
spurious contributions are the smallest when using a circle radius close to $R = 1$ for
the reasons outlined in Paper~II. Consequently, the corrected value is rather close to the uncorrected one in such a case.

It is fortuitous that for the deformation $Q_{20} = 500$ fm$^2$ in
\nuc{18}{O} and when using SLy4 the combined correction
of all density-independent terms is very small, such that corrected and
uncorrected curves are close at very small values of $R_p$ in
Fig.~\ref{shift} and even cannot be distinguished within the resolution of
the plot for larger $R_p$ shown.

Just as for the deformation energy curve as a function of quadrupole deformation, one
observes, by comparing the two panels of Fig.~\ref{shift}, that the steps generated
by SLy4 are significantly less steep than those produced by SIII before correction
(calculated in both cases with enough integration points to resolve them). This is due
to the fact that the steps are generated by a single pole leaving or entering the integration
circle in the case of SIII, which occurs over an infinitesimal variation of $R_{p}$,
whereas they are generated by a branch cut leaving or entering the integration
contour in the case of SLy4, which happens over a finite interval of variation of $R_{p}$.

Just as for the behavior of the deformation energy curve as a function of quadrupole
deformation, the curves obtained with 199 integration points in the upper panel of
Fig.~\ref{shift} show that the divergences seen before regularizing the quadratic
part of $\mathcal{E}^{\rho\rho\rho^{1/6}}$ have been replaced by cusps. Using more
mesh points for $R_{p}$ and $\varphi_{p}$, one could resolve even better the
non-derivable character of the PNR energy as the integration circle passes the
branch points. This is a direct illustration of the analytical result obtained
in Eq.~(\ref{derivee2}) and is schematically displayed in Fig.~\ref{pointanguleux}.

An important byproduct of the previous result is that they invalidate PNR calculations
performed using a fully antisymmetrized two-body interaction that depends on the medium
through a fractional power of the density, e.g.\ the Gogny interaction. The problem was
further circumvented in Ref.~\cite{anguiano01b} by using the projected density in place
of the transition density in the density-dependent term of the Gogny interaction. However,
such a procedure singles out one density factor in the energy kernel in a way that seems
highly arbitrary and not easily extendable to more involved EDFs. In addition, such a
prescription of using the correlated density into the density-dependent term of the
effective vertex leads to unsatisfactory results for other multi-reference calculations;
e.g.\ calculations including parity restoration and configuration mixing along the
octupole degree of freedom~\cite{robledo05a}.

\section{Summary and Conclusions}
\label{conclusions}

In Ref.~\cite{doba05a}, pathologies of calculations aiming at restoring particle
number and performed within the Energy Density Functional (EDF) framework have been
highlighted. In Ref.~\cite{lacroix06a}, the first paper of the present series, we
demonstrated that such pathologies are in fact shared by all multi-reference (MR)
calculations, i.e.\ symmetry restoration and/or Generator Coordinate Method (GCM)-based
configuration mixing calculations, performed within the EDF framework. In
Ref.~\cite{lacroix06a}, a formal and practical solution that applies (i) to any
symmetry restoration and/or GCM-based configuration mixing calculation (ii) to EDFs
depending only on integer powers of the density matrices, was proposed. In
Ref.~\cite{bender07x}, the second paper of the present series, the regularization
method was applied to Particle Number Restoration (PNR) calculations using an energy
functional that depends only on integer powers of the density matrices; e.g.\ which
contains linear, bilinear and trilinear terms.

The limitation of the correction method proposed in Ref.~\cite{lacroix06a} to energy
functionals depending on integer powers of the density matrices is a critical feature
as most functionals found in the literature contain non-integer powers of the (normal)
density matrix, both in the functional modeling the strong interaction and in
the functional modeling the Coulomb interaction, due to the Slater approximation to the
exchange term~\cite{bender03b}. Such non-integer powers of the density matrices pose
difficulties which go beyond those posed by integer powers: as transition densities
are complex, taking their non-integer powers amounts to dealing with a multivalued
function on the complex plane. This makes the analysis of the associated pathologies
more involved.

In the present paper, the third of the series, the viability of non-integer powers
of the density matrices has been addressed, building upon the analysis already carried
out in Ref.~\cite{doba05a}. First, we proposed to reduce the pathological character of
terms depending on a non-integer power of the density matrices by regularizing the
fraction that relates to the integer part of the exponent, using the method proposed in
Ref.~\cite{lacroix06a}. This amounts to scaling down the extent of the problem to the
one potentially encountered using a fully antisymmetrized effective interaction
depending further on a fractional power of the density; e.g.\ the Gogny force.
Second, we discussed in detail the spurious character of the remaining fractional
power of the density (matrix). Both through analytical derivations and numerical
applications (using the SLy4 Skyrme parametrization), we demonstrated that regularizing
the fraction related to the integer part of the exponent does remove divergences in the
particle number restored energy but replace them by cusps which are as unphysical as
the original divergences. In addition, the spurious steps in the PNR
energy and the related breaking of shift invariance prevail. Such results thus invalidate
PNR calculations performed using a fully antisymmetrized two-body interaction that depends
on the medium through a fractional power of the density.

Eventually, and because we do not see any well-defined basis to correct the corresponding
pathologies, we conclude at this point that non-integer powers of the density matrices are
not viable and should be avoided in the first place when constructing nuclear energy
density functionals to be used in MR-EDF calculations in the future. However, one will have to restrict the form to rather low integer orders in the density matrices. For example, the EDF recently proposed by Baldo \emph{et al}.\ \cite{baldo08a} includes terms up to fifth power in the total density $\rho(\vec{r})$, which lead to self-interaction terms~\cite{stringari78a} that will require a regularization containing quadruple sums over single-particle states, which will be too costly in realistic calculations.

Let us make an additional comment regarding the drastic
conclusion to discard non-integer powers of the density matrices altogether. On the one hand, integer powers of the density matrices appear naturally
when constructing the EDF through ab-initio calculations, e.g.\ through many-body
perturbation theory. On the other hand, non-integer powers of the density matrices,
if not introduced merely on phenomenological grounds, do often, if not always, result
from interpreting integrals over momenta up to $k_F$ providing the infinite matter equation
of state with contributions of the kind $k_F^n$ as density-dependent term through the use
of $k_F \sim \rho^{1/3}$. Transported to finite nuclei, where the latter relationship has no rigorous
basis, through some version of the local density approximation, this leads to an EDF that
contains terms of the form $\rho^{n/3}$. Although such a constructive procedure of the
nuclear EDF does not lead to particular problems in single reference (SR) calculations,
it does so when this procedure is extended to MR calculations as even the local part of the
scalar-isoscalar transition density matrix is complex, stretching one step too far the
above procedure proceeding through infinite matter and the use of
$k_F \Leftrightarrow \rho^{1/3}$. Finally, there are both practical reasons and formal
motivations to conclude that (i) non-integer powers of the density (matrix) are not
viable in (multi-reference) EDF calculations (ii) parameterizations making only use of
integer powers of the densities need to be constructed in the very near future. Last
but not least, note that such a conclusion actually extends to any form of the EDF that
generates branch cuts when continued over the complex.

\begin{acknowledgments}

This work was supported by the U.S.\ National Science Foundation under Grant No.\ PHY-0456903.
T.~L.\ and K.~B.\ thank the NSCL for the kind hospitality  during the completion of this work.

\end{acknowledgments}

\bibliography{papierIII}

\begin{thebibliography}{21}
\expandafter\ifx\csname natexlab\endcsname\relax\def\natexlab#1{#1}\fi
\expandafter\ifx\csname bibnamefont\endcsname\relax
  \def\bibnamefont#1{#1}\fi
\expandafter\ifx\csname bibfnamefont\endcsname\relax
  \def\bibfnamefont#1{#1}\fi
\expandafter\ifx\csname citenamefont\endcsname\relax
  \def\citenamefont#1{#1}\fi
\expandafter\ifx\csname url\endcsname\relax
  \def\url#1{\texttt{#1}}\fi
\expandafter\ifx\csname urlprefix\endcsname\relax\def\urlprefix{URL }\fi
\providecommand{\bibinfo}[2]{#2}
\providecommand{\eprint}[2][]{\url{#2}}

\bibitem[{\citenamefont{Dobaczewski et~al.}(2007)\citenamefont{Dobaczewski,
  Nazarewicz, Reinhard, and Stoitsov}}]{doba05a}
\bibinfo{author}{\bibfnamefont{J.}~\bibnamefont{Dobaczewski}},
  \bibinfo{author}{\bibfnamefont{W.}~\bibnamefont{Nazarewicz}},
  \bibinfo{author}{\bibfnamefont{P.~G.} \bibnamefont{Reinhard}},
  \bibnamefont{and} \bibinfo{author}{\bibfnamefont{M.~V.}
  \bibnamefont{Stoitsov}}, \bibinfo{journal}{Phys. Rev. C}
  \textbf{\bibinfo{volume}{76}}, \bibinfo{pages}{054315}
  (\bibinfo{year}{2007}).

\bibitem[{\citenamefont{D{\"o}nau}(1998)}]{donau98}
\bibinfo{author}{\bibfnamefont{F.}~\bibnamefont{D{\"o}nau}},
  \bibinfo{journal}{Phys. Rev. C} \textbf{\bibinfo{volume}{58}},
  \bibinfo{pages}{872} (\bibinfo{year}{1998}).

\bibitem[{\citenamefont{Anguiano et~al.}(2001)\citenamefont{Anguiano, Egido,
  and Robledo}}]{anguiano01b}
\bibinfo{author}{\bibfnamefont{M.}~\bibnamefont{Anguiano}},
  \bibinfo{author}{\bibfnamefont{J.~L.} \bibnamefont{Egido}}, \bibnamefont{and}
  \bibinfo{author}{\bibfnamefont{L.~M.} \bibnamefont{Robledo}},
  \bibinfo{journal}{Nucl. Phys.} \textbf{\bibinfo{volume}{A696}},
  \bibinfo{pages}{467} (\bibinfo{year}{2001}).

\bibitem[{\citenamefont{Almehed et~al.}(2001)\citenamefont{Almehed, Frauendorf,
  and D{\"o}nau}}]{almehed01a}
\bibinfo{author}{\bibfnamefont{D.}~\bibnamefont{Almehed}},
  \bibinfo{author}{\bibfnamefont{S.}~\bibnamefont{Frauendorf}},
  \bibnamefont{and}
  \bibinfo{author}{\bibfnamefont{F.}~\bibnamefont{D{\"o}nau}},
  \bibinfo{journal}{Phys. Rev.} \textbf{\bibinfo{volume}{C63}},
  \bibinfo{pages}{044311} (\bibinfo{year}{2001}).

\bibitem[{\citenamefont{Lacroix et~al.}(2008)\citenamefont{Lacroix, Duguet, and
  Bender}}]{lacroix06a}
\bibinfo{author}{\bibfnamefont{D.}~\bibnamefont{Lacroix}},
  \bibinfo{author}{\bibfnamefont{T.}~\bibnamefont{Duguet}}, \bibnamefont{and}
  \bibinfo{author}{\bibfnamefont{M.}~\bibnamefont{Bender}}
  (\bibinfo{year}{2008}), \bibinfo{note}{submitted to Phys. Rev. C},
  \eprint{arXiv:0809.2041}.

\bibitem[{\citenamefont{Bender et~al.}(2008)\citenamefont{Bender, Duguet, and
  Lacroix}}]{bender07x}
\bibinfo{author}{\bibfnamefont{M.}~\bibnamefont{Bender}},
  \bibinfo{author}{\bibfnamefont{T.}~\bibnamefont{Duguet}}, \bibnamefont{and}
  \bibinfo{author}{\bibfnamefont{D.}~\bibnamefont{Lacroix}}
  (\bibinfo{year}{2008}), \bibinfo{note}{submitted to Phys. Rev. C},
  \eprint{arXiv:0809.2045}.

\bibitem[{\citenamefont{Horen et~al.}(1996)\citenamefont{Horen, Satchler,
  Fayans, and Trykov}}]{horen96a}
\bibinfo{author}{\bibfnamefont{D.~J.} \bibnamefont{Horen}},
  \bibinfo{author}{\bibfnamefont{G.~R.} \bibnamefont{Satchler}},
  \bibinfo{author}{\bibfnamefont{S.~A.} \bibnamefont{Fayans}},
  \bibnamefont{and} \bibinfo{author}{\bibfnamefont{E.~L.}
  \bibnamefont{Trykov}}, \bibinfo{journal}{Nucl. Phys.}
  \textbf{\bibinfo{volume}{A600}}, \bibinfo{pages}{193} (\bibinfo{year}{1996}).

\bibitem[{\citenamefont{Fayans et~al.}(2000)\citenamefont{Fayans, Tolokonnikov,
  Trykov, and Zawischa}}]{fayans}
\bibinfo{author}{\bibfnamefont{S.~A.} \bibnamefont{Fayans}},
  \bibinfo{author}{\bibfnamefont{S.~V.} \bibnamefont{Tolokonnikov}},
  \bibinfo{author}{\bibfnamefont{E.~L.} \bibnamefont{Trykov}},
  \bibnamefont{and} \bibinfo{author}{\bibfnamefont{D.}~\bibnamefont{Zawischa}},
  \bibinfo{journal}{Nucl. Phys.} \textbf{\bibinfo{volume}{A676}},
  \bibinfo{pages}{49} (\bibinfo{year}{2000}).

\bibitem[{\citenamefont{Bender et~al.}(2003)\citenamefont{Bender, Heenen, and
  Reinhard}}]{bender03b}
\bibinfo{author}{\bibfnamefont{M.}~\bibnamefont{Bender}},
  \bibinfo{author}{\bibfnamefont{P.-H.} \bibnamefont{Heenen}},
  \bibnamefont{and} \bibinfo{author}{\bibfnamefont{P.-G.}
  \bibnamefont{Reinhard}}, \bibinfo{journal}{Rev. Mod. Phys.}
  \textbf{\bibinfo{volume}{75}}, \bibinfo{pages}{121} (\bibinfo{year}{2003}).

\bibitem[{\citenamefont{Dobaczewski and Dudek}(1995)}]{doba95a}
\bibinfo{author}{\bibfnamefont{J.}~\bibnamefont{Dobaczewski}} \bibnamefont{and}
  \bibinfo{author}{\bibfnamefont{J.}~\bibnamefont{Dudek}},
  \bibinfo{journal}{Phys. Rev.} \textbf{\bibinfo{volume}{C52}},
  \bibinfo{pages}{1827} (\bibinfo{year}{1995}).

\bibitem[{\citenamefont{Henley and Wilets}(1964)}]{henley}
\bibinfo{author}{\bibfnamefont{E.~M.} \bibnamefont{Henley}} \bibnamefont{and}
  \bibinfo{author}{\bibfnamefont{L.}~\bibnamefont{Wilets}},
  \bibinfo{journal}{Phys. Rev.} \textbf{\bibinfo{volume}{133}},
  \bibinfo{pages}{B1118} (\bibinfo{year}{1964}).

\bibitem[{\citenamefont{Bulgac}(2002)}]{bulgac1}
\bibinfo{author}{\bibfnamefont{A.}~\bibnamefont{Bulgac}},
  \bibinfo{journal}{Phys. Rev.} \textbf{\bibinfo{volume}{C65}},
  \bibinfo{pages}{051305} (\bibinfo{year}{2002}).

\bibitem[{\citenamefont{Robledo}(2007)}]{Robledo07a}
\bibinfo{author}{\bibfnamefont{L.~M.} \bibnamefont{Robledo}},
  \bibinfo{journal}{Int. J. Mod. Phys.} \textbf{\bibinfo{volume}{E16}},
  \bibinfo{pages}{337} (\bibinfo{year}{2007}).

\bibitem[{\citenamefont{Bayman}(1960)}]{Bay60a}
\bibinfo{author}{\bibfnamefont{B.~F.} \bibnamefont{Bayman}},
  \bibinfo{journal}{Nucl. Phys.} \textbf{\bibinfo{volume}{15}},
  \bibinfo{pages}{33} (\bibinfo{year}{1960}).

\bibitem[{\citenamefont{Duguet}(2006)}]{duguet06b}
\bibinfo{author}{\bibfnamefont{T.}~\bibnamefont{Duguet}}
  (\bibinfo{year}{2006}), \eprint{unpublished}.

\bibitem[{\citenamefont{Stoitsov et~al.}(2007)\citenamefont{Stoitsov,
  Dobaczewski, Kirchner, Nazarewicz, and Terasaki}}]{stoitsov07a}
\bibinfo{author}{\bibfnamefont{M.~V.} \bibnamefont{Stoitsov}},
  \bibinfo{author}{\bibfnamefont{J.}~\bibnamefont{Dobaczewski}},
  \bibinfo{author}{\bibfnamefont{R.}~\bibnamefont{Kirchner}},
  \bibinfo{author}{\bibfnamefont{W.}~\bibnamefont{Nazarewicz}},
  \bibnamefont{and} \bibinfo{author}{\bibfnamefont{J.}~\bibnamefont{Terasaki}},
  \bibinfo{journal}{Phys. Rev.} \textbf{\bibinfo{volume}{C76}},
  \bibinfo{pages}{014308} (\bibinfo{year}{2007}).

\bibitem[{\citenamefont{Bender and Heenen}(2003)}]{bender03c}
\bibinfo{author}{\bibfnamefont{M.}~\bibnamefont{Bender}} \bibnamefont{and}
  \bibinfo{author}{\bibfnamefont{P.-H.} \bibnamefont{Heenen}},
  \bibinfo{journal}{Nucl.\ Phys.} \textbf{\bibinfo{volume}{A713}},
  \bibinfo{pages}{390} (\bibinfo{year}{2003}).

\bibitem[{\citenamefont{Chabanat et~al.}(1998)\citenamefont{Chabanat, Bonche,
  Haensel, Meyer, and Schaeffer}}]{chabanat98}
\bibinfo{author}{\bibfnamefont{E.}~\bibnamefont{Chabanat}},
  \bibinfo{author}{\bibfnamefont{P.}~\bibnamefont{Bonche}},
  \bibinfo{author}{\bibfnamefont{P.}~\bibnamefont{Haensel}},
  \bibinfo{author}{\bibfnamefont{J.}~\bibnamefont{Meyer}}, \bibnamefont{and}
  \bibinfo{author}{\bibfnamefont{R.}~\bibnamefont{Schaeffer}},
  \bibinfo{journal}{Nucl. Phys.} \textbf{\bibinfo{volume}{A635}},
  \bibinfo{pages}{231} (\bibinfo{year}{1998}).

\bibitem[{\citenamefont{Robledo}(2005)}]{robledo05a}
\bibinfo{author}{\bibfnamefont{L.}~\bibnamefont{Robledo}}
  (\bibinfo{year}{2005}), \bibinfo{note}{beyond Mean Field Calculation With The
  Density Dependent Gogny Force, invited talk, INT workshop on Nuclear
  Structure Near the Limits of Stability, September 26 to December 2, 2005}.

\bibitem[{\citenamefont{Baldo et~al.}(2008)\citenamefont{Baldo, Schuck, and
  Vi{\~n}as}}]{baldo08a}
\bibinfo{author}{\bibfnamefont{M.}~\bibnamefont{Baldo}},
  \bibinfo{author}{\bibfnamefont{P.}~\bibnamefont{Schuck}}, \bibnamefont{and}
  \bibinfo{author}{\bibfnamefont{X.}~\bibnamefont{Vi{\~n}as}},
  \bibinfo{journal}{Phys. Lett.} \textbf{\bibinfo{volume}{B663}},
  \bibinfo{pages}{390} (\bibinfo{year}{2008}).

\bibitem[{\citenamefont{Stringari and Brink}(1978)}]{stringari78a}
\bibinfo{author}{\bibfnamefont{S.}~\bibnamefont{Stringari}} \bibnamefont{and}
  \bibinfo{author}{\bibfnamefont{D.~M.} \bibnamefont{Brink}},
  \bibinfo{journal}{Nucl. Phys.} \textbf{\bibinfo{volume}{A304}},
  \bibinfo{pages}{307} (\bibinfo{year}{1978}).

\end{thebibliography}

\end{document}